\RequirePackage{fix-cm}
\documentclass[twocolumn,natbib]{svjour3}
\smartqed  
\journalname{Journal of Computational Neuroscience}

\newcommand{\todothese}[1]{}

\usepackage[table, usenames, dvipsnames]{xcolor}

\usepackage{amsmath}

\usepackage{graphicx}
\usepackage{grffile}
\usepackage{float}
\usepackage{color}
\definecolor{MyDarkBlue}{rgb}{0,0.08,0.45}

\usepackage[utf8]{inputenc}
\usepackage[english]{babel}

\usepackage{boites,boites_exemples}

\usepackage{wasysym}

\usepackage{etoolbox}
\usepackage{pgf,pgffor}
\usepackage{tikz}
\usepackage{alphalph}

\usepackage{placeins}

\usepackage{ctable}

\newcommand{\fr}[3]{$#1 \pm #2 (#3)$}

\newcommand{\xMSN}[1]{#1_{\mbox{\tiny \ \scriptsize MSN}}}
\newcommand{\xFSI}[1]{#1_{\mbox{\tiny \ \scriptsize FSI}}}
\newcommand{\xSTN}[1]{#1_{\mbox{\tiny \ \scriptsize STN}}}
\newcommand{\xGPe}[1]{#1_{\mbox{\tiny \ \scriptsize GPe}}}
\newcommand{\xGPi}[1]{#1_{\mbox{\tiny \ \scriptsize GPi}}}
\newcommand{\xCMPf}[1]{#1_{\mbox{\tiny \ \scriptsize CM/Pf}}}
\newcommand{\xyMSN}[2]{#1^{\mbox{\tiny \ \scriptsize #2}}_{\mbox{\tiny \ \scriptsize MSN}}}
\newcommand{\xyFSI}[2]{#1^{\mbox{\tiny \ \scriptsize #2}}_{\mbox{\tiny \ \scriptsize FSI}}}
\newcommand{\xySTN}[2]{#1^{\mbox{\tiny \ \scriptsize #2}}_{\mbox{\tiny \ \scriptsize STN}}}
\newcommand{\xyGPe}[2]{#1^{\mbox{\tiny \ \scriptsize #2}}_{\mbox{\tiny \ \scriptsize GPe}}}
\newcommand{\xyGPi}[2]{#1^{\mbox{\tiny \ \scriptsize #2}}_{\mbox{\tiny \ \scriptsize GPi}}}

\newcommand{\xyz}[3]{#1^{\mbox{\tiny \ \scriptsize #2}}_{\mbox{\tiny \ \scriptsize #3}}}
\newcommand{\xampa}[1]{\mbox{#1}_{\mbox{\scriptsize AMPA}}}
\newcommand{\xnmda}[1]{\mbox{#1}_{\mbox{\scriptsize NMDA}}}
\newcommand{\xgabaa}[1]{\mbox{#1}_{\mbox{\scriptsize GABA}_{\mbox{\tiny A}}}}
\newcommand{\gabaa}{\mbox{GABA}_{\mbox{\footnotesize A}}}
\newcommand{\txtgabaa}{$\mbox{GABA}_{\mbox{\footnotesize A}}$}
\newcommand{\xynu}[2]{\nu_{\mbox{\scriptsize #2} \leftarrow \mbox{\scriptsize #1}}}
\newcommand{\xyalpha}[2]{\alpha_{\mbox{\scriptsize #1} \rightarrow \mbox{\scriptsize #2}}}
\newcommand{\xyp}[2]{p_{\mbox{\scriptsize #1} \rightarrow \mbox{\scriptsize #2}}}
\newcommand{\xyP}[2]{\mathcal{P}_{\mbox{\scriptsize #1} \rightarrow \mbox{\scriptsize #2}}}

\newcommand{\xyphi}[2]{\phi^{\mbox{\scriptsize #1}}_{\mbox{\scriptsize #2}}}
\newcommand{\xysigma}[2]{\sigma^{\mbox{\scriptsize #1}}_{\mbox{\scriptsize #2}}}

\newcounter{tnotenb}%
\newcommand\citable{}
\newcommand\listofnotes{}
\newcommand\tobeadded{}

\newcommand\cftxt{*}
\newcolumntype{R}[1]{>{\raggedleft\arraybackslash}p{#1}}
\newcommand\hl[1]{\bf #1}

\newcommand\myifstrequal{\expandafter\ifstrequal\expandafter}

\makeatletter

\newcommand{\notectable}[4]
{

\renewcommand\listofnotes{}
\renewcommand\tobeadded{}

\renewcommand{\citable}[1]%
{%
\renewcommand\tobeadded{##1}%
\setcounter{tnotenb}{0}%
\@for\tnotetext:=\listofnotes\do{%
\stepcounter{tnotenb}%
 \myifstrequal{\tnotetext}{##1}{\textsuperscript{\AlphAlph{\value{tnotenb}}}%
\renewcommand\tobeadded{}%
\global\csdef{txt\tnotetext}{\expandafter\tnotetext}%
}{}%
}%
\myifstrequal{\tobeadded}{}{}{%
\stepcounter{tnotenb}%
\myifstrequal{\listofnotes}{}{\gappto\listofnotes{##1}}{\gappto\listofnotes{,##1}}%
\textsuperscript{\AlphAlph{\value{tnotenb}}}%
\global\csdef{txt##1}{##1}%
}%
}

\ctable[%
doinside=\footnotesize,
#1%
]{#2}{%
\setcounter{tnotenb}{0}%
\@for\tnotetext:=\listofnotes\do{%
\stepcounter{tnotenb}%
 \textsuperscript{\AlphAlph{\value{tnotenb}}}:\cite{\csuse{txt\tnotetext}}\ \ %
}}{#3}%
}

\makeatother

\author{Jean Li{\'e}nard$^{1,2}$ \and Beno\^{i}t Girard$^{1,2}$}
\title{A Biologically Constrained Model of the Whole Basal Ganglia Addressing the Paradoxes of Connections and Selection.\thanks{This research was funded by the ANR, project EvoNeuro ANR-09-EMER-005-01.}
}

\institute{
Corresponding author: J. Liénard \at
Institut des Systèmes Intelligents et de Robotique\\
Université Pierre et Marie CURIE - Pyramide - T55/65\\
CC 173 - 4 Place Jussieu - 75005 Paris\\
\vspace{0.5cm}
\noindent\email{\{lienard,benoit.girard\}@isir.upmc.fr}             \\
{\scriptsize 1. UPMC Univ Paris 06, UMR 7222, ISIR, F-75005, Paris, France\\
2. CNRS, UMR 7222, ISIR, F-75005, Paris, France }
}

\begin{document}
\sloppy


\maketitle

\begin{abstract}

The basal ganglia nuclei form a complex network of nuclei often assumed to perform selection, yet their individual roles and how they influence each other is still largely unclear. In particular, the ties between the external and internal parts of the globus pallidus are paradoxical, as anatomical data suggest a potent inhibitory projection between them while electrophysiological recordings indicate that they have similar activities. Here we introduce a theoretical study that reconciles both views on the intra-pallidal projection, by providing a plausible characterization of the relationship between the external and internal globus pallidus. Specifically, we developed a mean-field model of the whole basal ganglia, whose parameterization is optimized to respect best a collection of numerous anatomical and electrophysiological data. We first obtained models respecting all our constraints, hence anatomical and electrophysiological data on the intrapallidal projection are globally consistent. This model furthermore predicts that both aforementioned views about the intra-pallidal projection may be reconciled when this projection is weakly inhibitory, thus making it possible to support similar neural activity in both nuclei and for the entire basal ganglia to select between actions. Second, we predicts that afferent projections are substantially unbalanced towards the external segment, as it receives the strongest excitation from STN and the weakest inhibition from the striatum. Finally, our study strongly suggest that the intrapallidal connection pattern is not focused but diffuse, as this latter pattern is more efficient for the overall selection performed in the basal ganglia.


\end{abstract}



\section{Introduction}


The basal ganglia, a fundamental component of the learning and decision processes of the vertebrate brain, comprises several nuclei which are interconnected through a complex connectivity pattern whose corresponding functionalities are not deciphered yet and still a subject of intense studies. We conducted a computational study through a model of the whole basal ganglia in primate, realized with population coding (with mean-field models) and integrating detailed anatomical data like the size of dendritic fields, the respective number of neurons in the nuclei and the number of axonal varicosities. Such a large model requires the adjustment of more than 50 parameters, which we achieved using stochastic optimization heavily constrained by experimental data.

The main aim of this computational study is to investigate the relations between the external and internal parts of the globus pallidus inside the whole basal ganglia (BG). One key issue in BG understanding is the conciliation of anatomical data indicating that the external part of the globus pallidus exerts a direct inhibitory influence on its internal part, with electrophysiological data showing that these nuclei have apparently similar activities \citep{Mink96,Nambu08}. By studying this connection both from the anatomical and from the functional point of view, we aim to provide a further understanding of its place and role in the BG.

The second aim is to study more globally the physiology of the whole primate BG, which differs by several points from the rodent. An up-to-date view of the BG connections graph show that, in primate, nearly all the possible connections are present (Figure \ref{fig:realisticbg}). Our goal is to provide a rigorous quantification of the strength of all the connections, that is consistent with as much data as possible from the literature.

The third and final aim is to contribute to the understanding of the function performed by the BG. The now classical theory that a selection is being performed in the BG \citep{Mink96,Redgrave99} has been illustrated by several computational models (e.g. \cite{Berns94,Gurney01a, Frank06, Leblois06, Humphries06, Girard08}). The behavior of these models is highly dependent of their parameterizations, which have been obtained through fine-tuning under various arbitrary assumptions. By exploring the search space of plausible model parameterizations while making as few simplifications as possible, we aim to determine the extent to which the selection theory is effectively supported by an extensive collection of data on the primate BG.

While we are using population level coding of the BG, our work aims at including detailed anatomical data that are usually used in modeling with spiking neurons, while still being able to respect electrophysiological data acquired in several monkey {\it in vivo} experiments. Moreover, our work aims to respect the different confidence ranges up to which the data are known. To do so, we formalize the modeling problem in terms of optimizing model fit to biological data, and to solve it we have developed an original methodology based on multi-objective optimization algorithms.

The objectives are dual in our approach. First, we want the design of our model to match detailed quantitative data concerning the anatomy and structure of the BG, in other words, we want it to be plausible {\it by construction}. Yet this objective is not sufficient by itself, as there is no telling how realistic a model will behave when simulated, no matter how good it looks on the paper. So, our second objective is that the model should {\it exhibit a behavior in simulation} as close as possible to the real BG. Considering that our level of modeling permits mainly to monitor firing rates of the nuclei, this second objective translates as the capability to fit to rates recorded in electrophysiological studies. These two objectives are similar to the {\it construct} and {\it face} validity concepts that have long been employed when discussing the validity of animal models in neuropathology \citep[e.g.][]{Willner84,Geyer95,Nestler10}, which explains why we dubbed our objectives with these terms.

Finally, our level of modeling rests on data whose availability varies widely among the species. Given the wide amount of axonal tracing studies done in the monkey, we choose this species to gather our data. When not specified, the references of this paper concerns the macaque, and we explicitly mention it when we cite works on other primate genus or on other species.

\section{Materials and Methods}

\subsection{Network Architecture of the BG}

The BG is composed of four different nuclei: the Striatum (Str), the Globus Pallidus, subdivided in an external part (GPe) and an internal part (GPi), the Subthalamic nucleus (STN) and the Substantia nigra, which is subdivided in a dopaminergic part (pars compacta, SNc) and a gabaergic part (pars reticulata, SNr). The BG receives inputs from the Cortex and the Parafascicular and Centromedian thalamic nuclei (CM/Pf).
In the classical view \citep{Alexander90,Mink96}, the Str and STN send projections to the GPe and GPi; the GPe sends projection to the GPi and STN; the Cortex sends projections to the Str and STN; the CM/Pf, finally, innervates the Str. The striatofugal projections are furthermore subdivided in a {\it direct} pathway targeting the GPi/SNr and expressing D1 Dopamine receptor, as well as an {\it indirect} pathway targeting the GPe and bearing D2 receptor.

\begin{figure}[h]
\centering
\includegraphics[width=0.4\textwidth]{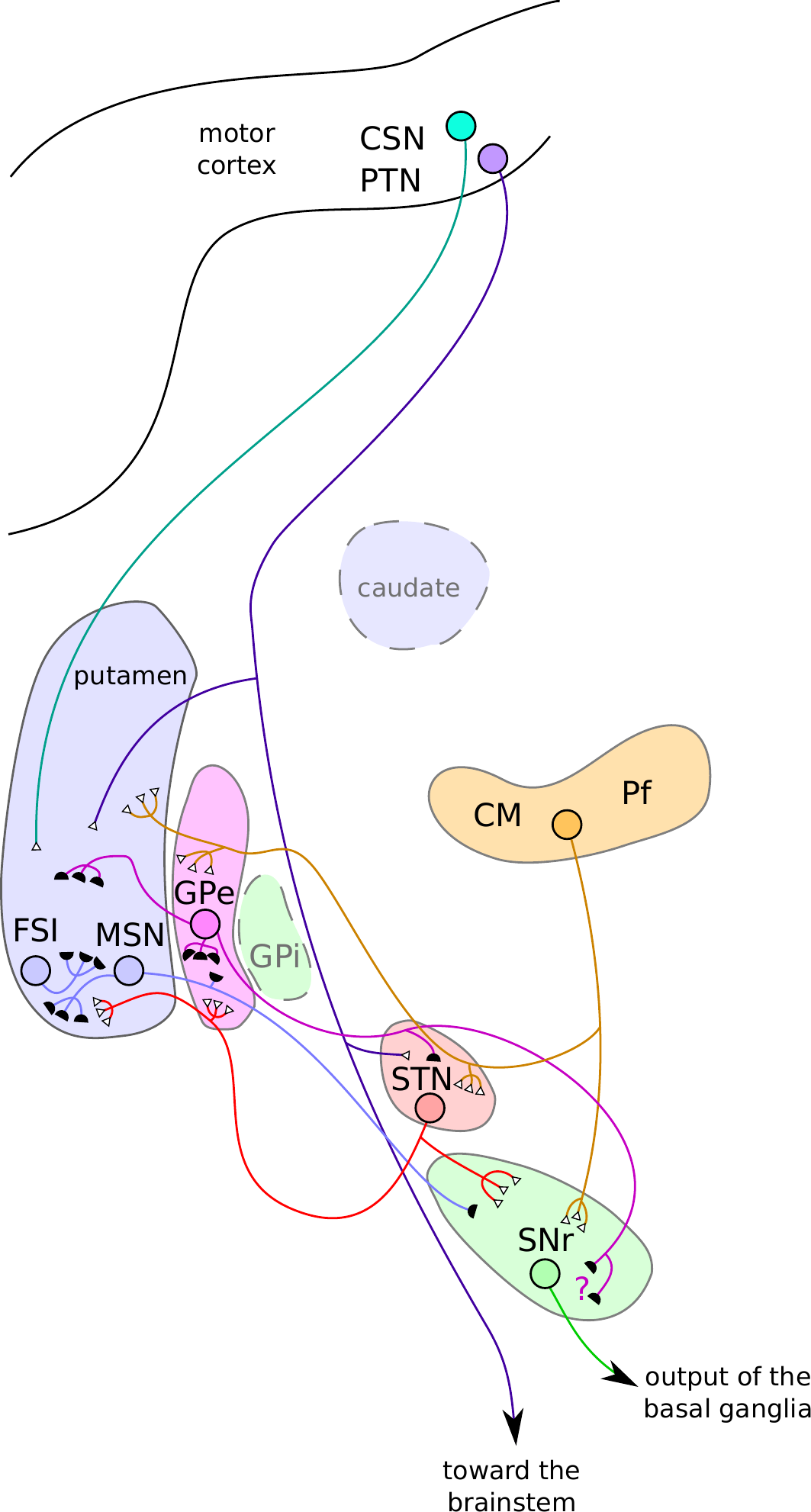}
\caption{Schematic illustration of the primate BG connectivity, for one cluster of neurons. White endings are excitatory, black are inhibitory. Multiple endings denote a diffuse connection, while a single ending indicate a focused connection. For the sake of clarity, only the SNr connections are represented, the GPi connections being similar; likewise, connections are represented for the putamen but not for the caudate. Projections ending in the striatum target with a similar pattern both MSN and FSI, apart from the MSN recurrent connection that does not target FSI. The outline of the BG structures was partly reproduced from \cite{Parent06} and \cite{Tande06}.}
\label{fig:realisticbg}
\end{figure}

While this view has been the common ground for a long time to understand the BG processing in both rodents and primates, several findings, both old and new, have called for a more detailed understanding of the primate BG.
In particular in primate, but also to a lesser extent in rodent, the simple view in which the direct and indirect pathways are perfectly segregated appears to be an over-simplification. Indeed, two different single axons tracing studies have shown that almost every MSN target both the GPe and GPi/SNr \citep{Parent95c,Levesque05a}, implying that most direct MSN must project to GPe and most indirect MSN must also project to GPi/SNr. Moreover, {\it in situ} hybridization studies of D1 and D2 receptors in the striatum have provided estimates of colocalization ranging from 5\% \citep{Aubert00} to 50\% \citep{Nadjar06}, consistent with the idea that these pathways are overlapping.
Also, some axons originating from the STN target the striatum, as several studies have reported through the years using different techniques \citep{Nauta78,Parent87,Nakano90,Smith90a,Sato00b}.
Furthermore, a projection from the GPe to the striatum has been reported in primate \citep{Beckstead83,Spooren96,Kita99,Sato00a}, although there is no study yet showing whether it targets both projection neurons and interneurons as in rodents \citep{Bevan98,Mallet12}.
Concerning the BG inputs, the centromedian complex of the intralaminar thalamus (CM/Pf) sends projections to the whole BG and not only to the striatum \citep{Sadikot92b,Parent05,Tande06}.
Finally, the cortical afferences can be subdivided in two pathways, with cortico-striatal neurons (CSN) projecting solely to the striatum and pyramidal tract neurons (PTN) targeting both the striatum and STN \citep{Parent06}.
See Figure \ref{fig:realisticbg} for an actualized view of the primate anatomy of the BG.

\subsection{Neuron population model}

\subsubsection{Mathematical formulation}

We use a population model with mean-field formulation. Although we provide here the basic equations of our model, more details about this formulation can be found elsewhere \citep{Deco08}.

One assumption of mean-field models, commonly referred as the {\it diffusion approximation}, is that every neuron receive the same inputs from another population. We can hence express the mean number of incoming spikes with neurotransmitter $n$ per neuron of the population $x$ from population $y$:

\begin{eqnarray}
\Psi_{x}^{n}(t) = \nu_{i \leftarrow y}\phi_k(t-\tau_{y \rightarrow x})
\end{eqnarray}

with 
$\nu_{x \leftarrow y}$ the mean number of synapse in one neuron of population $x$ from axons of population $y$, $\tau_{y \rightarrow x}$ the axonal delay between population $y$ and $x$, and $\phi_y(t-\tau_{y \rightarrow x})$ the firing rate of population $y$ at time $t-\tau_{y \rightarrow x}$.

In order to be able to express in population $x$ the mean count of synapses targeted by population $y$, we can rely on axonal varicosity counts:

\begin{eqnarray}
\label{equ:axon_to_synapse}
\nu_{x \leftarrow y} = \frac{\xyP{y}{x} N_y}{N_x}.\alpha_{y \rightarrow x}
\end{eqnarray}

with $N_x$ and $N_y$ the neuron counts of populations $x$ and $y$, $\alpha_{y \rightarrow x}$ the mean axonal varicosity count of neurons of $y$ with an axon targeting neurons of $x$, and $\xyP{y}{x}$ the proportion of such neurons in population $y$.

One more assumption of mean-field models is that the firing thresholds of the neurons of any population are distributed as a Gaussian function. The mean firing rate of a population $x$ at time $t$ can then be approximated by:

\begin{eqnarray}
\phi_x(t) = \frac{\xyz{S}{max}{x}}{1 + \mathrm{exp}(\frac{\theta_x-\Delta V_x(t)}{\sigma\prime})}
\end{eqnarray}

with $\Delta V_x(t)$ the mean input potential at the soma at time $t$, $\xyz{S}{max}{x}$ the maximal possible firing rate, $\theta_x$ the mean difference between resting and firing thresholds, and, as per \cite{VanAlbada09a}, $\sigma\prime = \sigma \frac{\sqrt{3}}{\pi}$ with $\sigma$ the standard deviation of the firing thresholds.

We use so-called alpha functions to describe the post-synaptic change caused by a single spike mediated by a neurotransmitter $n$, to the membrane potential at the location of the synapse:
\begin{eqnarray}
V_{0}^{n}(t) = A_n D_n t e^{-D_n t}
\end{eqnarray}

with $A_n$ the amplitude and $D_n$ the half-life, which are values specific to each neurotransmitter $n$. 

We also model in a simple way the attenuation of the distal dendrites as a function of the soma distance. By modeling the dendritic field as a single compartment finite cable with sealed-end boundaries condition \citep{Koch05}, we can express for population $x$:

\begin{eqnarray}
\label{equ:dendrite}
  V_{\mbox{\scriptsize soma}}^{n}(t) = V_{0}^{n}(t)\ \frac{cosh(L_x-Z_x)}{cosh(L_x)}
\end{eqnarray}

with $V_{0}^{n}(t)$ the potential change at the synapse, $L_x$ the electrotonic constant of the neurons and $Z_x$ the mean distance of the synaptic receptors along the dendrites. We further express this mean distance as a percentage of the mean maximal dendritic length $l_x$: $Z_x = p_x L_x $. The electrotonic constant is then calculated according to \citep{Koch05}:

\begin{eqnarray}
\label{equ:electrotonic_constant}
  L_x = l_x\ \sqrt{\frac{4}{d_x}\frac{R_i}{R_m}\ }
\end{eqnarray}

with $R_i$ the intracellular resistivity, $R_m$ the membrane resistance, $l_x$ the mean maximal dendritic length and $d_x$ the mean diameters of the dendrites along their whole extent for population $x$.

Finally, we can express $\Delta V_x(t)$ the mean change of potential of one population caused by all its afferents:

\begin{eqnarray}
\label{equ:sum_aff}
  \Delta V_x(t) = \sum_{(y,n)} \Psi_{x}^{n}(t) V_{\mbox{\scriptsize soma}}^{n}(t)
\end{eqnarray}

with each couple $(y,n)$ representing one afferent population $y$ whose spikes are mediated by a neurotransmitter $n$.

\subsubsection{Global constants}

In order to use the population model described in the previous part, we need to provide some global parameters which are not specific to the BG populations. Most of these parameters are unknown for the primate nervous system, therefore also unknown for the primate BG, because of the lack of single cell electrophysiological studies. 

In the absence of data on the firing threshold spread,
we opted for the same plausible value $\sigma \approx 7 \mbox{ mV}$ as in \cite{VanAlbada09a}, resulting in $\sigma' = 3.8 \mbox{ mV}$. Membrane resistance and intracellular resistivity, which are used to calculate the attenuation as a function of synapse distance, have been chosen to standard values with $R_m = 20000\ \Omega.\mbox{cm}^2$ and $R_i = 200\ \Omega.\mbox{cm}$ \citep{Koch05}.
The post-synaptic potential (PSP) half-times were directly derived from the study of \cite{Destexhe98a} (see Table \ref{tab:mf_fixed_parameters}).
Following the arguments exposed in \cite{Rodrigues10} on the relative strength of AMPA over $\gabaa$ unitary PSP, we take their ratio to be 4:1 and set $\xampa{A} = 1 \mbox{ mV}$, $\xgabaa{A} = 0.25 \mbox{ mV}$.
Indeed, these values permit to recreate several estimates from electrophysiological studies in the rodent striatum, where minimal values of inhibitory PSP are in the range 0.17-0.34 mV \citep[c.f. the review of][]{Tepper04b}, and AMPA excitatory PSP have been shown to be a little less than 1mV \citep{Carter07}.
The ratio of AMPA over NMDA contribution has been taken to be 2, as the deactivations studies that we aim to reproduce also show that inhibition by NBQX is roughly two times more effective than inhibition by CPP \citep{Kita04,Tachibana08}. This is done by setting $\xnmda{A} = 0.025 \mbox{ mV}$ while keeping the decay time in accordance with \cite{Destexhe98a}.

\notectable{
botcap,
caption = {Mean-field constants. Asterix denote that the detailed justification is to be found in the text.},%
label = {tab:mf_fixed_parameters},notespar}
{lrp{1cm}rp{0.3cm}}
{ 
\hl{Parameter} & \hl{Unit} & \hl{Symbol} & \hl{Value} & \hl{Ref} \ML
Neuronal properties & & & & \\   \cmidrule(r){1-2}
  Threshold spread & (mV) & $\sigma'$ & 3.8 & \citable{VanAlbada09a}\\
  Membrane resistance & ($\Omega.\mbox{cm}^{2}$) & $R_m$ & 20000 & \citable{Koch05}\\
  Intracellular resistivity & ($\Omega.\mbox{cm}$) & $R_i$ & 200 & \citable{Koch05}\\
\\
PSP amplitudes & (mV) & & & \\   \cmidrule(r){1-2}
  AMPA & & $\xampa{A}$ & 1 & \cftxt\\
  $\gabaa$ & & $\xgabaa{A}$ & 0.25 & \cftxt\\
  NMDA & & $\xnmda{A}$ & 0.025 & \cftxt\\
\\
PSP half-times & (ms) & & & \\   \cmidrule(r){1-2}
  AMPA & & $\xampa{D}$ & 5 & \citable{Destexhe98a}\\
  $\gabaa$ & & $\xgabaa{D}$ & 5 & \citable{Destexhe98a}\\
  NMDA & & $\xnmda{D}$ & 100 &  \citable{Destexhe98a}
\ML}

\subsection{Parameters : Fixed vs Optimized}

When constructing a quantitative modeling of a neuronal circuit of the size and complexity of the BG, it is rather difficult to attribute a value for each parameter. In our approach, we subdivided the parameters in two groups. The first group comprises (1) the parameters that are pretty well known and (2) parameters for which a slight change around their suspected value would not lead to a meaningful interpretation in the context of this study. The second group comprises the opposite parameters, i.e. (1) the ones whose plausible order of magnitude is broad, and (2) those whose study will likely meets our aims.
The parameters in this second group are optimized in order to fit best to their plausible range.

\subsubsection{Fixed Parameters}


The neuron numbers, the diameters and maximal extents of the dendrites, the maximal discharge rates (except for the FSI), the axonal delays and the proportion of projection neurons, belong to the group of fixed parameters. See Table \ref{tab:fixed_parameters} for an exhaustive list of their values.

The counts of the STN, GPe and GPi/SNr neurons come straightforwardly from the stereotaxic study of \cite{Hardman02}. The count of matrix MSN ({\it resp.} FSI) was obtained by combining their total number estimated at 30 400 000 ({\it resp.} 611 000) in the patch and matrix \citep{Yelnik91} with the percentage of matrix neurons, which accounts for 87\% of the whole striatum \citep{Johnston90}. No direct count of the CM/Pf projection neurons in the macaque has been found, so the neuron count of 86 000 has been deduced by the combining the volume of $28.35 \mbox{ mm}^{3}$ for CM ({\it resp.} $14.28 \mbox{ mm}^{3}$ for Pf) \citep{Paxinos00}, the mean neuronal density of 2759 neurons.mm$^{-3}$ for CM ({\it resp.} 3655 neurons.mm$^{-3}$) and the proportion of glutamatergic neurons of $67\%$ for CM ({\it resp.} $66\%$) as these are the efferent projection neurons \citep{Hunt91}.

At odds with rodents, there are no {\it in vitro} studies of the properties of the primate BG neurons, so these have to be taken from {\it in vivo} studies. The MSN have recently been reported to be able to fire at a frequency that can reach 300 Hz \citep{Nambu09a}. \cite{Matsumura92,Nambu00,Iwamuro07} reported that the STN could reach 300 Hz, while \cite{Wichmann06} reported peak intraburst frequencies of $214.0 \pm 85.2$ Hz, so we set the maximal firing rate of the STN at 300 Hz. The Globus Pallidus, both external \citep{Nambu00,Kita04} and internal \citep{Nambu00,Tachibana08}, can reach a firing rate of 400 Hz, and \cite{Wichmann06} reported peak intraburst frequencies of $300.4 \pm 97.3$ Hz and $322.7 \pm 72.7$ Hz, so we set the maximal firing rate of both of them at 400 Hz.

The firing rates at rest are set to 2 Hz for the cortico-striatal neurons (CSN) and 15 Hz for the pyramidal tract neurons (PTN) of the cortex \citep{Bauswein89,Turner00}, as well as 4 Hz for the CM/Pf \citep{Matsumoto01}.

The dendrites morphology being approximated as a cylinder (c.f. equation \ref{equ:dendrite}), we gathered informations from morphological studies to determine their mean maximal extents and mean diameters. The mean maximal dendritic extents of both MSN and FSI were computed from the data of \cite{Yelnik91} using the formula $Ln.S/F$, with $Ln$ the mean length of all dendritic segments, $S$ the number of stems and $F$ the number of tips. This resulted for MSN ({\it resp.} FSI) in an estimate of the mean maximal extent of 619 $\mu$m ({\it resp.} 961 $\mu$m), while their mean diameters along the dendritic tree has been estimated to be 1 $\mu$m for the MSN and 1.5 $\mu$m for the FSI. The STN data were gathered from the anatomical studies of \cite{Rafols76,Yelnik79} which show that 750 $\mu$m is an adequate estimate for the maximal extent. Upon examination of these two studies, 1.5 $\mu$m was chosen to reflect the mean value of dendrite diameters. A detailed model of the GPe ({\it resp.} GPi) neurons in the macaque \citep{Mouchet04} permitted us to estimate the mean maximal extent 1132 $\mu$m ({\it resp.} 865 $\mu$m ) using the formula $Ln.S/F$ (see their Table I at p. 14) as well as the mean diameter of 1.2 $\mu$m ({\it resp.} 1.7 $\mu$m).

Finally, single cell tracing studies in the primate have revealed that the projection neurons of a nucleus are not uniformly sending their axon toward a target nucleus. Indeed, \cite{Levesque05a} quantified that 82\% of the striatal projection neurons target GPi/SNr, while all of them target GPe. In another study, \cite{Sato00b} reported that 83\% of STN neurons target GPe, 72\% target GPi/SNr and 17\% could be followed until the striatal border, most probably to target striatal neurons. In \cite{Sato00a}, 84\% of GPe target both STN and GPi/SNr while 16\% target striatum. While we are aware that these proportions have been established on limited and possibly biased samples, we included them as fixed parameters because these are ultimately combined with the bouton counts (c.f. equation \ref{equ:axon_to_synapse}) which are optimized in a broad range in order to compensate for this.

\notectable{
botcap,
caption = {Fixed parameters of the BG. The projections for which 100\% of the afferent neurons display axonal varicosities in the target nucleus (e.g. MSN~$\rightarrow$~GPi/SNr) are not reported in this table.},%
label = {tab:fixed_parameters},notespar}
{p{3.375cm}p{0.6cm}p{1cm}rp{0.3cm}}
{ 
\hl{Parameter} & \hl{Unit} & \hl{Symbol} & \hl{Value} & \hl{Ref} \ML
Neurons number & ($10^{3}$) & & &  \\  \cmidrule(r){1-2}
%
  MSN & & $\xMSN{n}$ & 26 448 & \citable{Johnston90}\citable{Yelnik91} \\
  FSI & & $\xFSI{n}$ & 532 & \citable{Johnston90}\citable{Yelnik91} \\
  STN & & $\xSTN{n}$ & 77 & \citable{Hardman02} \\
  GPe & & $\xGPe{n}$ & 251 & \citable{Hardman02} \\
  GPi/SNr & & $\xGPi{n}$ & 143 & \citable{Hardman02} \\
  CM/Pf (projection) & & $\xCMPf{n}$ & 86 & \cftxt \\
\\
Maximal firing rate & (Hz) & & & \\   \cmidrule(r){1-2}
  MSN & & $ \xyMSN{S}{max}$ & 300 & \citable{Nambu09a} \\
  STN & & $ \xySTN{S}{max}$ & 300 & \citable{Matsumura92}\citable{Nambu00}\citable{Wichmann06}\citable{Iwamuro07} \\
  GPe & & $ \xyGPe{S}{max}$ & 400 & \citable{Nambu00}\citable{Kita04}\citable{Wichmann06} \\
  GPi/SNr & & $ \xyGPi{S}{max}$ & 400 & \citable{Nambu00}\citable{Wichmann06}\citable{Tachibana08} \\
\\
Firing rate at rest & (Hz) & & & \\   \cmidrule(r){1-2}
   CSN & & $ \xyphi{}{CSN} $ & 2 & \citable{Bauswein89}\citable{Turner00} \\
   PTN & & $ \xyphi{}{PTN} $ & 15 & \citable{Bauswein89}\citable{Turner00} \\
   CM/Pf & & $ \xyphi{}{CMPf} $ & 4 & \citable{Matsumoto01} \\
\\
Mean dendritic extent & ($\mu$m) & & & \\   \cmidrule(r){1-2}
  MSN & & $ \xyMSN{l}{max} $ & 619  & \citable{Yelnik91} \\
  FSI & & $ \xyFSI{l}{max} $ & 961  & \citable{Yelnik91} \\
  STN & & $ \xySTN{l}{max} $ & 750  & \citable{Rafols76}\citable{Yelnik79} \\
  GPe & & $ \xyGPe{l}{max} $ & 1132 & \citable{Mouchet04} \\
  GPi/SNr & & $ \xyGPi{l}{max} $ & 865  & \citable{Mouchet04} \\
\\
Mean dendritic diameter & ($\mu$m) & & & \\   \cmidrule(r){1-2}
  MSN & & $ \xyMSN{d}{max} $ & 1 & \citable{Yelnik91} \\
  FSI & & $ \xyFSI{d}{max} $ & 1.5 & \citable{Yelnik91} \\
  STN & & $ \xySTN{d}{max} $ & 1.5 & \citable{Yelnik79} \\
  GPe & & $ \xyGPe{d}{max} $ & 1.2 & \citable{Mouchet04} \\
  GPi/SNr & & $ \xyGPi{d}{max} $ & 1.7 & \citable{Mouchet04} \\
\\
\multicolumn{2}{l}{\% of projection neurons}  & & & \\   \cmidrule(r){1-2}
  MSN~$\rightarrow$~GPe & & $\xyP{MSN}{GPe}$ & 82\% & \citable{Levesque05a} \\
  STN~$\rightarrow$~GPe & & $\xyP{STN}{GPe}$ & 83\% & \citable{Sato00b} \\
  STN~$\rightarrow$~GPi/SNr & & $\xyP{STN}{GPi}$ & 72\% & \citable{Sato00b} \\
  STN~$\rightarrow$~MSN & & $\xyP{STN}{MSN}$ & 17\% & \citable{Sato00b} \\
  STN~$\rightarrow$~FSI & & $\xyP{STN}{FSI}$ & 17\% & \citable{Sato00b} \\
  GPe~$\rightarrow$~GPe & & $\xyP{GPe}{GPe}$ & 84\% & \citable{Sato00a} \\
  GPe~$\rightarrow$~GPi/SNr & & $\xyP{GPe}{GPi}$ & 84\% & \citable{Sato00a} \\
  GPe~$\rightarrow$~MSN & & $\xyP{GPe}{MSN}$ & 16\% & \citable{Sato00a} \\
  GPe~$\rightarrow$~FSI & & $\xyP{GPe}{FSI}$ & 16\% & \citable{Sato00a}
%
\ML}

We do not develop the temporal aspect in this work less apart for a check of the plausibility of the selection ability; for the sake of completeness, we used the temporal delays given in \cite{VanAlbada09a} and used 3 ms for the striatal afference from the GPe and STN, yet any change in the delays do not alter the results presented here.

\subsubsection{Optimized Parameters}


The optimized parameters are the mean number of synapses receiving cortical efferents in the striatum and STN, the mean axonal varicosity count of each other connection, the mean distance to soma of the receptors along the dendrites for each connection, the difference between the resting and threshold potential for each population and the maximal firing rate of the FSI. See Table \ref{tab:evolution_parameter_ranges} for an exhaustive list of their authorized boundaries.

An axon can have up to a few thousand varicosities, each one being a privileged area to establish a synaptic contact with a dendrite. In order to be able to express all the axonal types of the BG, we allowed each axonal varicosity count to take any value in the range of $0-6000$.

This strategy is not possible for the corticofugal neurons projecting to the BG because, while single axon tracing studies have been done in the monkey \citep{Parent06}, an evaluation of the cortico-basal neurons number remains an hazardous task. Yet synaptic studies at their targets (i.e. MSN, FSI and STN) indicate that the mean number of synapses arising from the cortex fall within the same wide range of $0-6000$ \citep{Yelnik79,Yelnik91}, allowing us to use it for the $\xynu{CSN}{{\it x }}$ and $\xynu{PTN}{{\it x }}$ parameters.

Another parameter used in the modeling is the mean distance of the synapses to the soma. In order to be able to differently reflect the influence of proximal and distal synapses, we optimized the distance to the soma as a percentage, with 0\% corresponding to synapses located at the soma and 100\% at the end of the dendritic tree.

While setting the maximal firing rates of most populations was quite straightforward, we could not find any study indicating what value it could take for the FSI. In vitro studies in the rodents indicate that this value is approximately 300 Hz in vitro \citep{Plenz98,Plotkin05,Gittis10} and a peak intraburst rate of 500 Hz has been recorded in vivo \citep{Berke04} and in vitro \citep{Kita91}, but we have no clue whether this is the same in the primate. In order to be on a comparable order of magnitude, we hence allowed $\xyFSI{S}{max}$ to evolve in the 200-500 Hz range.

Finally, the mean-field formulation uses the difference between resting potentials and threshold potentials, which are not known for the monkey BG. Even when looking at other species data, it is unclear what should be this difference, e.g. \cite{RavAcha05} would suggest a value in the 5-15 mV range for the MSN but one could deduce from \cite{Nambu94} a value more or less in the 15-30 mV range (this holds true for all the neuronal populations considered). In the absence of primate data and of other species clear consensual data, we cannot set these parameters and allow their evolution in the 5-30 mV range for each population.

\notectable{
botcap,
caption = {Allowed parameters boundaries.},%
label = {tab:evolution_parameter_ranges},notespar}
{lrR{1cm}r}
{ 
\hl{Parameter} & \hl{Unit} & \hl{Symbol} & \hl{Range} \ML
%
Corticofugal synapse count & & $\xynu{CSN}{{\it x }}$ $\xynu{PTN}{{\it x }}$  & 0-6000 \\
Axonal varicosity count & & $\xyalpha{{\it x}}{{\it y}}$ & 0-6000 \\
Synapse location & (\%) & $p_{\it x}$ & 0-100 \\
Relative firing thresholds & (mV)  & $\theta_{x}$ & 5-30 \\
Maximal FSI firing rate & (Hz) & $\xyFSI{S}{max}$ & 200-500
\ML}

\subsection{Objective functions of the optimization}

As stated in introduction, we elaborated two quantitative objectives to assess the fit of the models with available anatomical data, and with {\it in vivo} electrophysiological experiments.

\subsubsection{Error function}

To evaluate whether any parameter $p$ fitted well with the plausible range $[m,M]$, we attributed a score with an error function $e(p,m,M)$. To this end, we used the continuous juxtaposition of a Gaussian distribution outside the plausible range, and the constant function 1 inside, which is formally defined as:

\begin{eqnarray*}
e(p,m,M) = \left\{
  \begin{array}{ll}
		1 & \mbox{if } p \in [m,M]\\
		exp\left(- \frac{ 2(m-p)(M-p) }{ (m-M)^{2}  }\right) & \mbox{else}
  \end{array}
\right.
\end{eqnarray*}

\begin{figure}[tpb]
\centering
\includegraphics[width=0.45\textwidth]{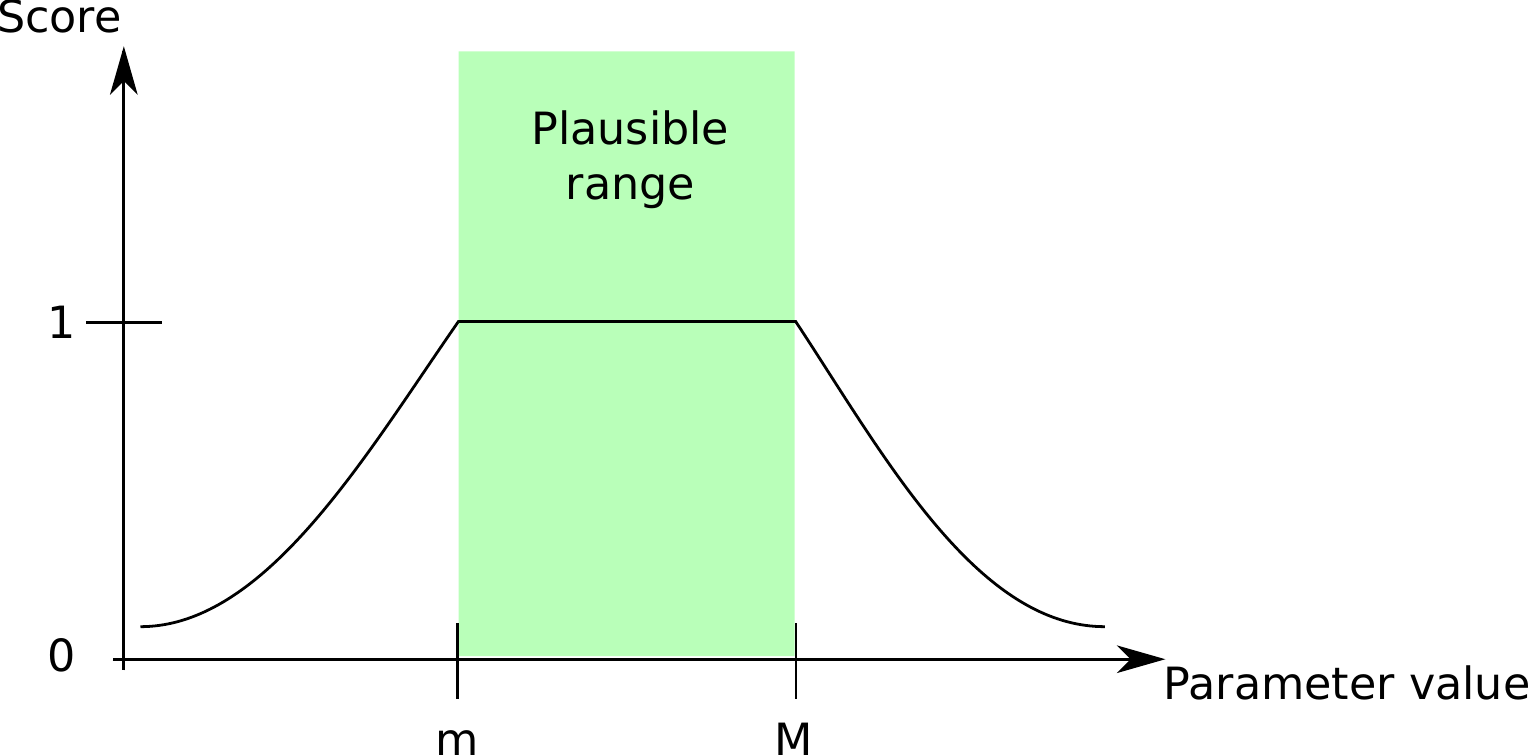}
\caption{Illustration of the error function $f$.}
\label{fig:plausible_range}
\end{figure}

This function is illustrated in Figure \ref{fig:plausible_range}. Note that while different function could be defined to represent the distance error between a parameter and its boundaries, the basic rectangular function valued 1 inside the plausible range and 0 outside would not be a good choice for optimization purpose. Indeed, it would not allow to distinguish between different distances outside the plausible range, and as consequence the adaptation process of the evolution would be made more difficult.

\subsubsection{Construct objective}

As previously stated, the first objective is used to assess whether the model is plausible {\it by construction}. There are 18 connection strengths evaluated on two levels: the mean number of varicosities per axon ({\it alt.} synapses per dendritic tree), and the mean distance between the synapses and the soma. Setting aside the few parameters for which we could not estimate a plausible range due to the lack of available data, there are 31 parameters to be checked against the plausible ranges given on Table \ref{tab:optimized_parameters}.

We divided the plausible ranges for each connection after careful consideration of the single axon tracing study available in the primate BG. Three subdivisions were established: (1) a low range, with axonal varicosity number inferior to 150; (2) a high range between 150 and 750;  (3) a very high range between 1000 and 5000. Upon collection of as much data from published single axon tracing studies as possible, we were able to categorize the parameters $\xyalpha{x}{y}$ as shown in Table \ref{tab:optimized_parameters}. One exception was the STN~$\rightarrow$~striatum connection for which no single axon tracing study was found, so it was evaluated on the basis of other imaging studies \citep{Nakano90,Nauta78,Parent87,Sato00a,Smith90a,Parent95b}. Another exception was the GPe~$\rightarrow$~GPi/SNr connection, for which no count was available. By looking at the single axon tracing study of \cite{Sato00b} and comparing to the other connection counts, we can deduce that it could either be in the low or high range, thus we set its plausible range to be between 15 and 750 boutons.

The synaptic counts matching CSN axonal varicosities was difficult to assess, and this overlaps a general difficulty to conciliate purely anatomical data with electrophysiological data. Indeed, the primate anatomical study of \cite{Yelnik91} showed that the number of asymmetric synapses, hence of cortical or thalamic origin, should be less than 4000-5000 (this upper bound is deduced from the maximal density of 8 synapses per 10 $\mu$m and the mean dendritic length of 5756 $\mu$m). However, after a careful review of the existing literature for the rodent, \cite{Humphries09} conclude that most of these synapses are inactive, and settled for a count of 250 active cortico-striatal synapses (see their section 3.2 p.1179). In order to reflect this disparity, we set the plausible range to 250-5000 CSN synapses. In the absence of data for FSI, we adopted the same very broad range for the CSN~$\rightarrow$~FSI projection.

The PTN axonal connectivity was also difficult to assess. As the striatal afference from the cortex is thought to be originating mostly from CSN \citep{Parent06} and as a previous study failed to activate antidromically PTN with a stimulation in the striatum \citep{Turner00}, the plausible ranges for the synaptic count of PTN~$\rightarrow$~MSN and PTN~$\rightarrow$~FSI connections was set from 1 to 1000, so there is a possibility to set this connection to negligible values during the optimization. The plausible range for the PTN~$\rightarrow$~STN connection was set to the widest range of all as we are aware of no data to constrain it; we considered that any value from 25 to 5000 synapses from PTN was plausible.


Finally, it must be emphasized that the maximal firing rate $\xyFSI{S}{max}$ and the difference between resting and firing potential $\theta$ for all populations are optimized but not scored by the construct objective, because plausible boundaries for these parameters could hardly be drawn. Hence they can take any value inside their evolution range without affecting the construct score.

\notectable{
botcap,
caption = {Anatomical objective.},%
label = {tab:optimized_parameters},notespar}
{p{3cm}rp{1.5cm}rp{0.35cm}}
{ 
\hl{Parameter} & \hl{Symbol} & \hl{Range} & \hl{Ref} \ML
Boutons number & & & \\   \cmidrule(r){1-1}
  MSN~$\rightarrow$~MSN & $\xyalpha{MSN}{MSN}$ & (150,750) & \cftxt \\
  FSI~$\rightarrow$~MSN & $\xyalpha{FSI}{MSN}$ & (1000,5000) & \cftxt \\
  FSI~$\rightarrow$~FSI & $\xyalpha{FSI}{FSI}$ & (15,150) & \cftxt \\
  MSN~$\rightarrow$~GPe & $\xyalpha{MSN}{GPe}$ & (150,750) & \citable{Levesque05b}\citable{Parent95c} \\
  MSN~$\rightarrow$~GPi/SNr & $\xyalpha{MSN}{GPi}$ & (150,750) & \citable{Levesque05a}\citable{Parent95c} \\
  STN~$\rightarrow$~GPe & $\xyalpha{STN}{GPe}$ & (150,750) & \citable{Parent07}\citable{Sato00b} \\
  STN~$\rightarrow$~GPi/SNr & $\xyalpha{STN}{GPi}$ & (150,750) & \citable{Parent07}\citable{Sato00b} \\
  STN~$\rightarrow$~striatum & $\xyalpha{STN}{Str}$ & (15,150) & \cftxt \\
  GPe~$\rightarrow$~GPe & $\xyalpha{GPe}{GPe}$ & (15,150) & \citable{Sato00a} \\
  GPe~$\rightarrow$~GPi/SNr & $\xyalpha{GPe}{GPi}$ & (15,750) & \citable{Sato00a} \\
  GPe~$\rightarrow$~STN & $\xyalpha{GPe}{STN}$ & (150,750) & \citable{Parent95b}\citable{Sato00a} \\
  CM/Pf~$\rightarrow$~striatum & $\xyalpha{Th}{Str}$ & (1000,5000) & \citable{Parent05} \\
  CM/Pf~$\rightarrow$~STN & $\xyalpha{Th}{STN}$ & (15,150) & \citable{Tande06} \\
  CM/Pf~$\rightarrow$~GPe & $\xyalpha{Th}{GPe}$ & (15,150) & \citable{Tande06} \\
  CM/Pf~$\rightarrow$~GPi/SNr & $\xyalpha{Th}{GPi}$ & (1,150) & \citable{Tande06} \\
\\
Synapses number  & & & \\   \cmidrule(r){1-1}
  MSN $\leftarrow$ CSN  & $\xynu{CSN}{MSN}$ & (250,5000) & \cftxt \\
  FSI $\leftarrow$ CSN  & $\xynu{CSN}{FSI}$ & (100,2500) & \cftxt \\
  MSN $\leftarrow$ PTN  & $\xynu{CSN}{MSN}$ & (1,1000) & \cftxt \\
  FSI $\leftarrow$ PTN  & $\xynu{PTN}{FSI}$ & (1,1000) & \cftxt \\
  STN $\leftarrow$ PTN  & $\xynu{PTN}{STN}$ & (25,5000) & \cftxt \\
\\
Receptor~location~(\%)   & & & \\   \cmidrule(r){1-1}
  cortex~$\rightarrow$~MSN  & $\xyp{CTX}{MSN}$ & (60,100) & \citable{Lapper92} \\
  cortex~$\rightarrow$~FSI  & $\xyp{CTX}{FSI}$ & (60,100) & \citable{Lapper92} \\
  cortex~$\rightarrow$~STN  & $\xyp{CTX}{STN}$ & (60,100) & \citable{Marani08a} \\
  MSN~$\rightarrow$~MSN  & $\xyp{MSN}{MSN}$ & (60,80) & \citable{Wilson07}\citable{Tepper08} \\ 
  MSN~$\rightarrow$~GPe  & $\xyp{MSN}{GPe}$ & (20,60) & \citable{Shink95}\citable{Shink96} \\
  MSN~$\rightarrow$~GPi/SNr  & $\xyp{MSN}{GPi}$ & (0,20) & \citable{Shink95}\citable{Shink96} \\
  FSI~$\rightarrow$~MSN  & $\xyp{FSI}{MSN}$ & (0,20) & \citable{Wilson07}\citable{Tepper08} \\ 
  STN~$\rightarrow$~GPe  & $\xyp{STN}{GPe}$ & (20,60) & \citable{Shink95}\citable{Shink96} \\
  STN~$\rightarrow$~GPi/SNr  & $\xyp{STN}{GPi}$ & (20,60) & \citable{Shink95} \\
  GPe~$\rightarrow$~STN  & $\xyp{GPe}{STN}$ & (20,60) & \citable{Shink96}\citable{Parent95b} \\
  GPe~$\rightarrow$~GPe  & $\xyp{GPe}{GPe}$ & (0,20) & \citable{Shink95} \\
  GPe~$\rightarrow$~GPi/SNr  & $\xyp{GPe}{GPi}$ & (0,20) & \citable{Shink95} \\
  CM/Pf~$\rightarrow$~MSN  & $\xyp{Th}{MSN}$ & (20,60) & \citable{Sidibe96}\citable{Sidibe99} \\
  CM/Pf~$\rightarrow$~FSI  & $\xyp{Th}{FSI}$ & (0,20) & \citable{Sidibe99}
\ML}

For each model, the construct objective is finally defined as a sum of error functions of the parameters $p_1,..,p_{31}$, given the plausible ranges $(m_1,M_1),..,(m_{31},M_{31})$ from Table \ref{tab:optimized_parameters}:

\begin{eqnarray*}
o_1 = \sum\limits_{i=1..31} e(p_i,m_i,M_i)
\end{eqnarray*}

\subsubsection{Face Value objective}

The second objective aims to assess whether the BG {\it activity} appears to be plausible. As this level of modeling permits us to, we formalized this objective as assessing whether the interactions between the different nuclei of the BG result in the right firing rates.

To do this, we assess first that the optimized model can reproduce the firing rates of the five neuronal populations at rest, then that it can reproduce the firing rates in nine pharmacological experiments of neurotransmitter antagonist injections (c.f. Table \ref{tab:firing_rates}).

Different difficulties arise in the elaboration of precise estimates for the mean firing rates at rest. In accordance with the general criterion, we checked whether the MSN firing rate at rest is less than 1 Hz. No electrophysiological study have been done that target specifically FSI in the monkey, but we can obtain some general orders for this rate from primate studies recording non-specified populations of the monkey striatum, giving mean overall firing rates of less than 10 Hz \citep{Aldridge90,Goldberg02}. However, these recordings could be an underestimation as studies recording rodents FSI in vivo indicate higher values \citep{Berke12}, hence we set the plausible range to the broad interval (0-20) Hz as a permissive compromise. On the contrary, a lot of electrophysiological studies are available on the monkey STN, GPe and GPi/SNr, and provide for each nucleus the firing rate and standard deviation of the studied sample whose size is also reported.
We included data from published recordings of some twenty monkeys and established mean estimates of 19.0 $\pm$ 3.8 for the STN, 65.1 $\pm$ 9.4 for the GPe and 69.3 $\pm$ 10.2 for the GPi/SNr (see appendix \ref{app:firing_rates_rest} for references and method).

We also scored the face value of the models with the ability to exhibit the same firing rates when a neurotransmitter antagonist is injected in the vicinity of the neurons, rendering the formers insensitive to the corresponding neurotransmitter. The data are taken from the two studies \cite{Kita04} and \cite{Tachibana08}, which used the antagonists CPP, NBQX and Gabazine, respectively inhibiting the effect of NMDA, AMPA and \txtgabaa. We used data for all the local injections they performed in GPe and GPi, resulting in nine firing rates changes. Following these articles, we supply all but one data in the form of a relative change. The only exception is the firing rate corresponding to the cumulative injections in GPi of all antagonists (i.e. CPP, Gabazine and NBQX) which was originally reported as an absolute value of 75.1 Hz $\pm$ 21.7\%. Note also that in the two cases named here GPe(2) and GPi(2) the firing rate is relative to the previous experiment. The firing rates at rest as well as with antagonist injections are summarized in Table \ref{tab:firing_rates}.

In order to reproduce the antagonist injections, we modeled each of them as a total deactivation of the concerned neurotransmitter. We hence simulate each model in the absence of antagonist, as well as one time for each of the nine antagonist conditions, giving ten sets of firing rates including the control for all nuclei $\xyphi{si}{MSN}, ... , \xyphi{si}{GPi}$ and the nine deactivation firing rates of GPe and GPi $\xyphi{si}{GPe(1)}, ... , \xyphi{si}{GPe(4)}, \xyphi{si}{GPi(1)}, ... , \xyphi{si}{GPe(5)}$.


For each model, the second objective is finally defined as the sum of the error functions of the simulated firing rates:

\begin{eqnarray*}
  \lefteqn{o_2 = }
	& & \ \ \ \ \ \ \sum\limits_{n=\{MSN, ..., GPi\}} e\left( \xyphi{si}{n},\xyphi{xp}{n}-\xysigma{xp}{n},\xyphi{xp}{n}+\xysigma{xp}{n}\right) \\
	& & + \sum\limits_{i=1..4} e\left( \xyphi{si}{GPe(i)}, \xyphi{si}{GPe} (1+\frac{\xyphi{xp}{GPe(i)}-\xysigma{xp}{GPe(i)}}{100}), \right. \\
	& & \ \ \ \ \ \ \ \ \ \ \ \ \ \ \left. \xyphi{si}{GPe} (1+\frac{\xyphi{xp}{GPe(i)}+\xysigma{xp}{GPe(i)}}{100})\right) \\
	& & + \sum\limits_{i=1..4} e\left( \xyphi{si}{GPi(i)}, \xyphi{si}{GPi} (1+\frac{\xyphi{xp}{GPi(i)}-\xysigma{xp}{GPi(i)}}{100}), \right. \\
	& & \ \ \ \ \ \ \ \ \ \ \ \ \ \ \left. \xyphi{si}{GPi} (1+\frac{\xyphi{xp}{GPi(i)}+\xysigma{xp}{GPi(i)}}{100})\right) \\
	& & +e\left( \xyphi{si}{GPi(5)},\xyphi{xp}{GPi(5)}-\xysigma{xp}{GPi(5)},\xyphi{xp}{GPi(5)}+\xysigma{xp}{GPi(5)}\right)
\end{eqnarray*}


\notectable{
botcap,
star,
width=\textwidth,
caption = {Electrophysiological objective.},%
label = {tab:firing_rates},notespar}{p{7cm}rp{4cm}rl}{ 
\hl{Parameter} & \hl{Unit} & \hl{Symbols} & \hl{Range} & \hl{Ref} \ML
Firing rates at rest & (Hz) & & & \cftxt \\  \cmidrule(r){1-2}
  \multicolumn{2}{l}{MSN}      & $\xyphi{xp}{MSN} \pm \xysigma{xp}{MSN}$ &  0.5 $\pm$  0.5 & \\
  \multicolumn{2}{l}{FSI}      & $\xyphi{xp}{FSI} \pm \xysigma{xp}{FSI}$ & 10   $\pm$ 10   & \\
  \multicolumn{2}{l}{STN}      & $\xyphi{xp}{STN} \pm \xysigma{xp}{STN}$ & 19.0 $\pm$  3.8 & \\
  \multicolumn{2}{l}{GPe}      & $\xyphi{xp}{GPe} \pm \xysigma{xp}{GPe}$ & 65.1 $\pm$  9.4 & \\
  \multicolumn{2}{l}{GPi/SNr}  & $\xyphi{xp}{GPi} \pm \xysigma{xp}{GPi}$ & 69.3 $\pm$ 10.2 & \\
%
\\
\% of rate changes in GPe &  & & & \citable{Kita04} \\  \cmidrule(r){1-2}
%
  \multicolumn{2}{l}{AMPA blocked}                             & $\xyphi{xp}{GPe(1)} \pm \xysigma{xp}{GPe(1)}$ & -  56.7 $\pm$ 35.6 \% &  \\
  \multicolumn{2}{l}{\txtgabaa blocked (after AMPA blockade)}  & $\xyphi{xp}{GPe(2)} \pm \xysigma{xp}{GPe(2)}$ & + 116.5 $\pm$ 16.7 \% &  \\
  \multicolumn{2}{l}{NMDA blocked}                             & $\xyphi{xp}{GPe(3)} \pm \xysigma{xp}{GPe(3)}$ & -  32.4 $\pm$ 14.5 \% &  \\
  \multicolumn{2}{l}{\txtgabaa blocked}                        & $\xyphi{xp}{GPe(4)} \pm \xysigma{xp}{GPe(4)}$ & + 115.8 $\pm$ 81.5 \% &  \\
%
\\
\% of rate changes in GPi &  & & & \citable{Tachibana08} \\  \cmidrule(r){1-2}
  \multicolumn{2}{l}{NMDA blocked}                       & $\xyphi{xp}{GPi(1)} \pm \xysigma{xp}{GPi(1)} $ & - 27.5 $\pm$  26.4 \% &  \\
  \multicolumn{2}{l}{AMPA blocked (after NMDA blockade)} & $\xyphi{xp}{GPi(2)} \pm \xysigma{xp}{GPi(2)} $ & - 54.2 $\pm$  20.8 \% &  \\
  \multicolumn{2}{l}{AMPA blocked}                       & $\xyphi{xp}{GPi(3)} \pm \xysigma{xp}{GPi(3)} $ & - 53.6 $\pm$  36.7 \% &  \\
  \multicolumn{2}{l}{\txtgabaa blocked}                  & $\xyphi{xp}{GPi(4)} \pm \xysigma{xp}{GPi(4)} $ & + 92.0 $\pm$ 117.3 \% &  \\
  \multicolumn{2}{l}{AMPA, NMDA and \txtgabaa blocked}   & $\xyphi{xp}{GPi(5)} \pm \xysigma{xp}{GPi(5)} $ &   75.1 $\pm$  21.7 \% &
\LL}

%

\subsection{Optimization algorithm}

We used the NSGA-2 evolutionary algorithm \citep{Deb02,mouret2010sferes} to optimize both objectives. We simulated each set of parameters for 15 seconds with a time step of $10^{-4}$ s, and discarded the set if it did not converge to fixed firing rates during this period. Each run made the evolution of the solutions for 1500 generations, with probabilities for mutation $\mu=0.1$ and cross-over $s=0.1$, and a population of 400 individuals. 100 evolution runs were launched at first. We furthermore enhanced the obtained results by performing a local search around the best solutions of this first pass, this was done with relaunching 100 optimization runs with an additional objective of diversity, and including all the solutions every 100 generations. The diversity was computed as the norm 1 difference between each solution and a given optimal solution from the previous runs.

In order to avoid duplicates and nearly-identical solutions among the final pool of solutions, we drew randomly couples of solutions whose normalized norm 1 difference was less than 1\permil \ and eliminated one of them chosen at random, until the minimal norm 1 difference between all the solution was more than 1\permil \ .


\subsection{Selection test}

\subsubsection{Subdivision in clusters}

In order to test whether the optimized models were capable of doing selection, we subdivided each nucleus into competitive channels, each corresponding to a different cluster of neurons putatively dedicated to a specific aspect of the information processing. This supposition has been commonly done in computational studies, it takes ground in the parallel loops organization of the whole primate BG \citep{Alexander86} and, for the particular case of motor control, in the somatotopic organization which occurs in all nuclei \citep{Nambu11}. The size of the clusters dedicated to each channel is unclear, in this work, we only supposed they contained enough neurons so that mean-field assumptions could still be true, without risking any hazardous estimate as we don't need an absolute number for those clusters to be modeled.

\begin{figure}[H]
\centering
\hfill \includegraphics[width=0.15\textwidth]{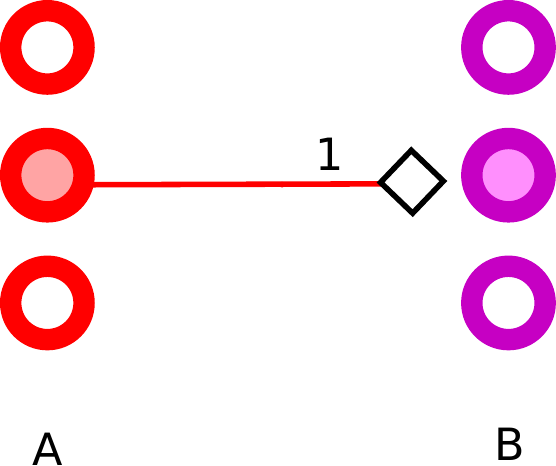} \hfill \includegraphics[width=0.15\textwidth]{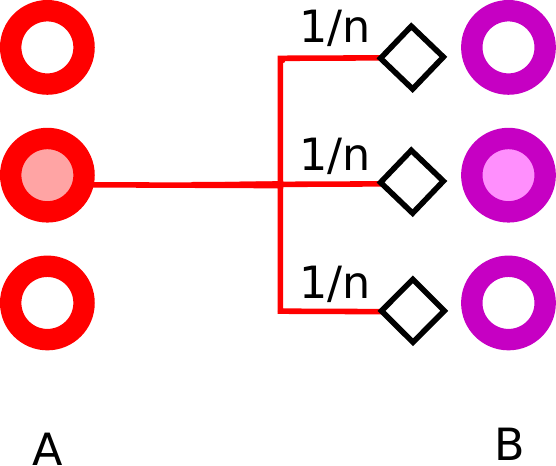} \hfill\
\caption{Illustration of the focused (left) and diffuse (right) connection types, with the indication of the strength coefficient with $n$ clusters.}
\label{fig:channels}
\end{figure}

We assumed that each connection belongs to one of two different connectivity schemes: either focused or diffuse. In the focused case, an axon from a given cluster from an initial nucleus A give all its varicosities in the corresponding cluster of a target B, while in the diffuse case an axon from a cluster of A gives an equal quantity of varicosities in every clusters of B (Figure \ref{fig:channels}).

In setting the connectivity type, we first supposed as a general rule that all local collaterals give diffuse connection, as they target neighboring neurons in the same nucleus. We also set the following outgoing connections to be diffuse:
FSI~$\rightarrow$~MSN, as the FSI individual activities are though to be synchronized through gap-junctions \citep{Koos99};
CM/Pf~$\rightarrow$~{\it all}, following \cite{Sadikot92a,Parent05} for the striatum and \cite{Sadikot92b,Tande06} for the GPe, STN and GPi/SNr;
GPe~$\rightarrow$~MSN and GPe~$\rightarrow$~FSI as per \cite{Spooren96};
STN~$\rightarrow$~{\it all}, based on \cite{Sato00b} for GPe and GPi/SNr and on \cite{Smith90a} for the striatum (see in particular their Figures 4 and 5).

We reciprocally considered the following outgoing connections as focused:
MSN~$\rightarrow$~GPe and GPi/SNr, as per the general consensus, which can be grounded by, e.g. \cite{Parent95c};
GPe~$\rightarrow$~STN following \citep{Sato00a};
all the cortical efferences, following the tracings by \cite{Parent06} and as our model supposes the same number of clusters in the cortex and in the BG, which means that the convergence of information at the cortico-striatal level is already included.

The only connection that we could not unambiguously set was the GPe~$\rightarrow$~GPi/SNr connection. The tracing study from \cite{Sato00a} could indicate any type of connectivity, so we tested both the focused and diffuse connectivity. See Figure \ref{fig:realisticbg} for a recapitulation of the connection patterns.

\subsubsection{Task}

The task was chosen to mimic a classical choice setup in monkey, which consists in making arm movements from the center and in eight directions at 45\textsuperscript{o} intervals, in a two dimensional apparatus. This task has been widely investigated, and the directionality of a subset of neurons at every stage of the cortico-basal circuit is well established. We explored here the part in which this task is already learned, so the cortico-striatal connections are set accordingly and no significant dopamine induced plasticity is thought to occur.

We set the inputs $\xyphi{i}{CSN}$ of each cluster to mimic the recorded activity in motor cortex as a sine wave maximizing a preferred direction at 0\textsuperscript{o} \citep{Georgopoulos82,Kalaska89}, with:

$$
\xyphi{i}{CSN} = \xyphi{}{CSN} (1 + sin(2i + 180)
$$

in which $i$ is a direction in the range [-180\textsuperscript{o}, +180\textsuperscript{o}] discretized in steps of 45\textsuperscript{o}. As we do not have a precise description of the temporal course of the cortical signal sent to the BG, we submit those constant entries until stabilization in the BG, at which point we examine the firing rates.

We used constant rest inputs in the PTN even though their values will likely change during the movement, because we are interested here in the BG main function of processing of the cortico-striatal inputs via the striatum. Indeed, PTN neurons are activated during motor control but the selection ability of the BG is thought not to be restricted to motor control. Likewise, we used constant rest inputs in the CM/Pf as we ignore if its activity is sensitive to directional cues, and even if it were, the principal information gate would still be the CSN from the whole cortex.

\subsection{Result analysis}

\subsubsection{Evaluation of selection}

The BG is thought to perform a selection by inhibition \citep{Mink96,Redgrave99}, in which the output activity of one cluster decreases as its input augments. In order to investigate the quality of the performed selection, we quantified it by calculating for each direction $i$ a contrast enhancement value $c_i$ as the normalized firing rate decrease of the GPi/SNr output compared to the corresponding increase of the cortico-striatal input:

$$
c_i = \frac{\xyphi{}{CSN}}{\xyphi{i}{CSN}} \times \frac{\xyphi{}{GPi}}{\xyphi{i}{GPi}}
$$

in which $\xyphi{}{CSN}$ is the baseline activity at 2Hz and $\xyphi{}{GPi}$ is the baseline activity of the GPi.
This provides an estimate of the contrast performed in output of each channel, normalized with respect to the channel inputs through the term $\xyphi{}{CSN}/\xyphi{i}{CSN}$. In the case of selection by inhibition, we expect $c_i$ to be low for deselected channels (with $\xyphi{i}{GPi} > \xyphi{}{GPi}$) and high for selected channels (with $\xyphi{i}{GPi} < \xyphi{}{GPi}$). The channel-wise difference of this indicator between two different connectivity patterns will be used to compare their impact on selection.

\subsubsection{Evaluation of the connection strength}

As stated before, two combined parameters set the connection strength of nucleus $x$ toward nucleus $y$: the average number of axonal varicosities $\alpha_{x\rightarrow y}$ and the average localization of the receptors along the dendrites $p_{x\rightarrow y}$. Different combinations of these parameters yield the same connection strength because the increase of one can be compensated by a decrease of the other, resulting in what we term the {\it isoforce} $f_{x\rightarrow y}$. The isoforce is formally defined as:

$$
f_{x\rightarrow y} =  \int_0^\infty \! \nu_{y\leftarrow x} V_{\mbox{\scriptsize soma}}^n(t) \, dt
$$

The value of $f_{x\rightarrow y}$ is dependent on both $\alpha_{x\leftarrow y}$ (used to compute $\nu_{y\leftarrow x}$) and $p_{x\rightarrow y}$ (used to compute $V_{\mbox{\scriptsize soma}}^n$). The above formula takes also into account the relative numbers of neurons in each nucleus as well as the dendritic constants of the targeted nucleus. The unit of $f_{x\rightarrow y}$ is $\mu V.s$, and its physiological interpretation is to link the incoming firing rate $\phi_x$ (in Hz) to the corresponding mean change of the potential $\Delta V_y$ (in $\mu V$) during one second, with
$f_{x\rightarrow y} \phi_x = \Delta V_y$
.

\section{Results}

\subsection{Optimized solutions}

We obtain 1468 optimal solutions which maximize both the construct and face objectives after optimization, meaning that all the plausibility constraints we put on the anatomy and electrophysiology were fully satisfied. Each solution corresponds to a different set of parameters, or in other words, each solution is a different model of the BG. Among this pool, we retain only 1151 solutions that are not duplicates (see methods) to lower the computational time of the foregoing analysis of the results.

\begin{figure}[htpb]
\centering
\includegraphics[width=0.4\textwidth]{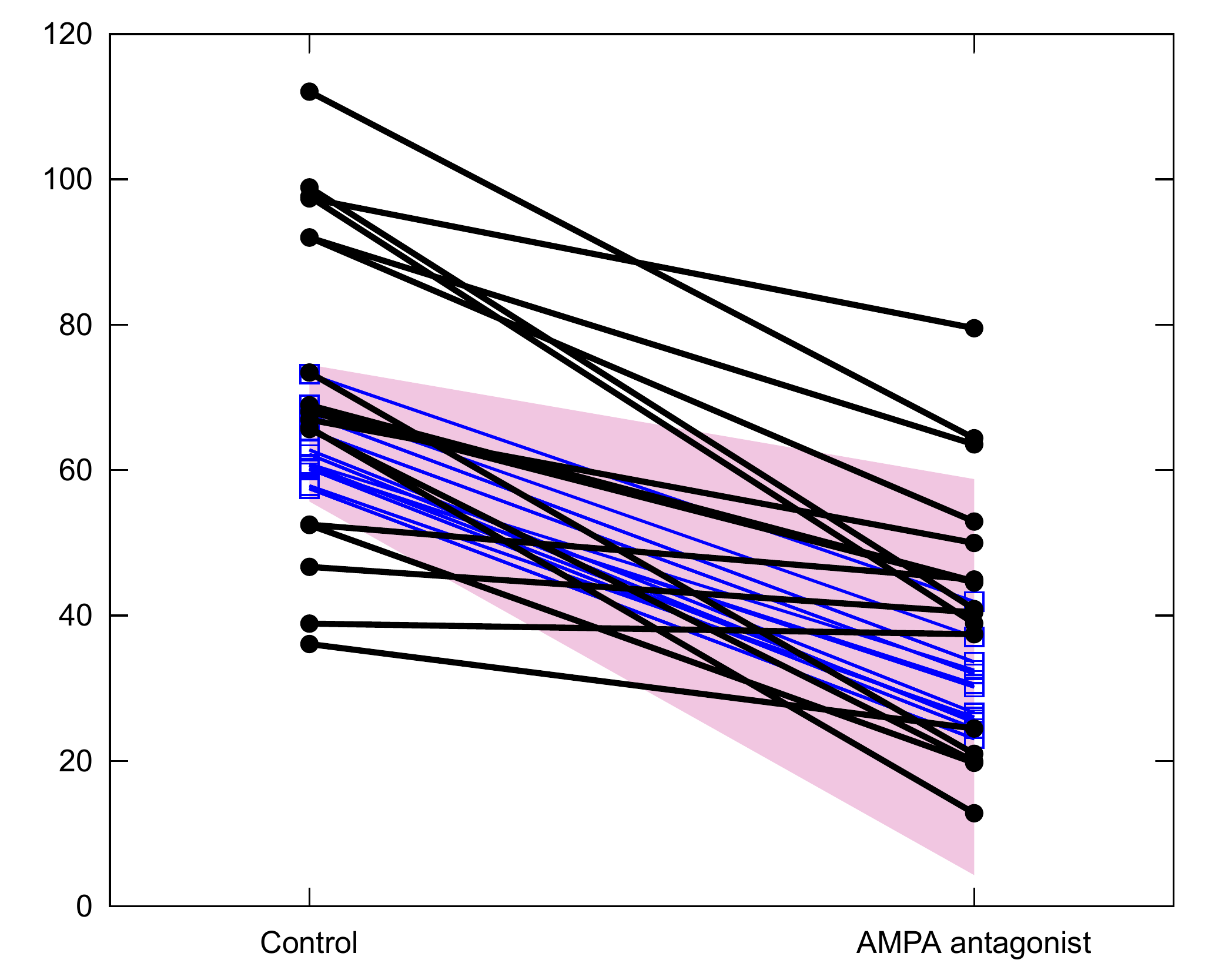}
\includegraphics[width=0.4\textwidth]{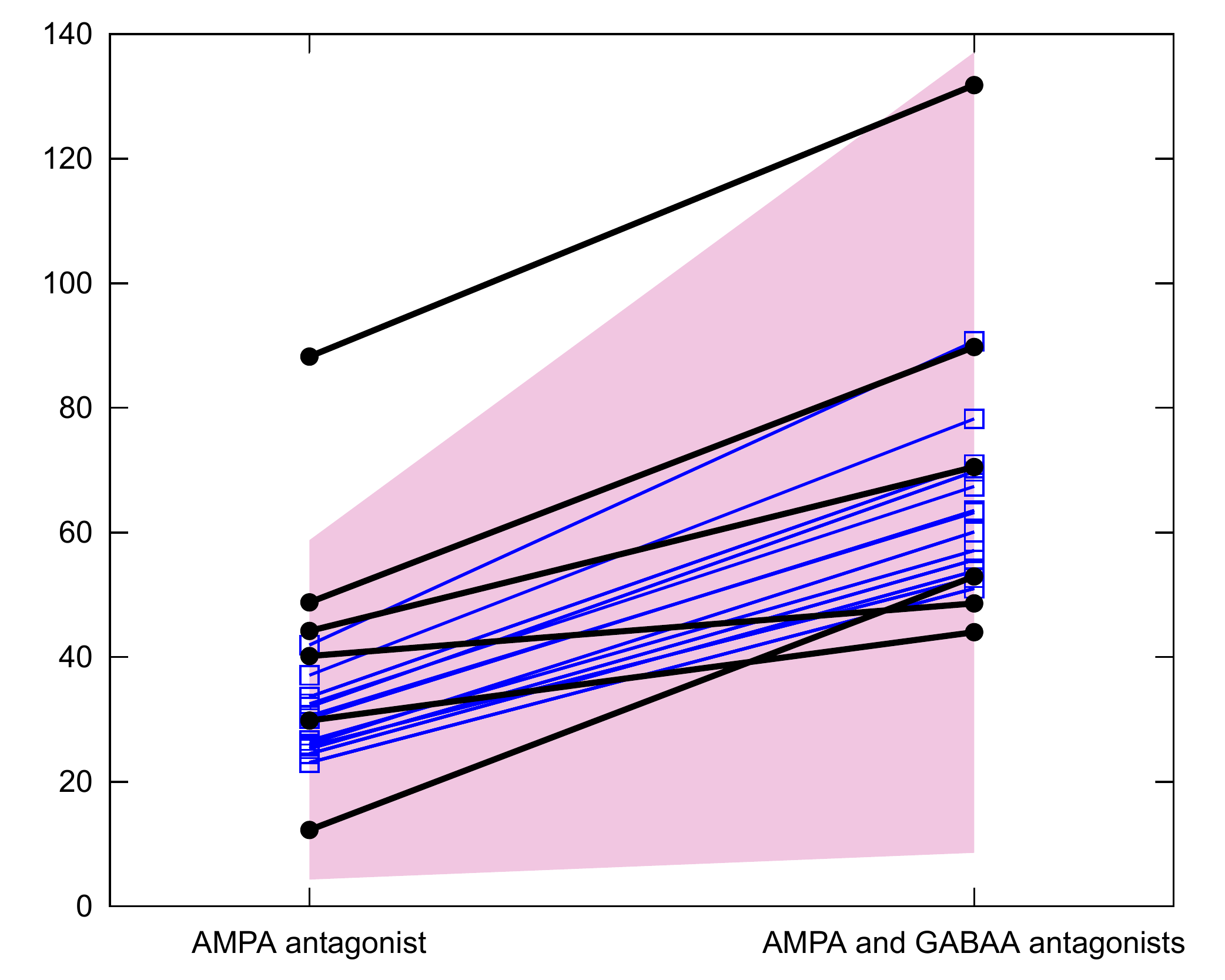}
\includegraphics[width=0.4\textwidth]{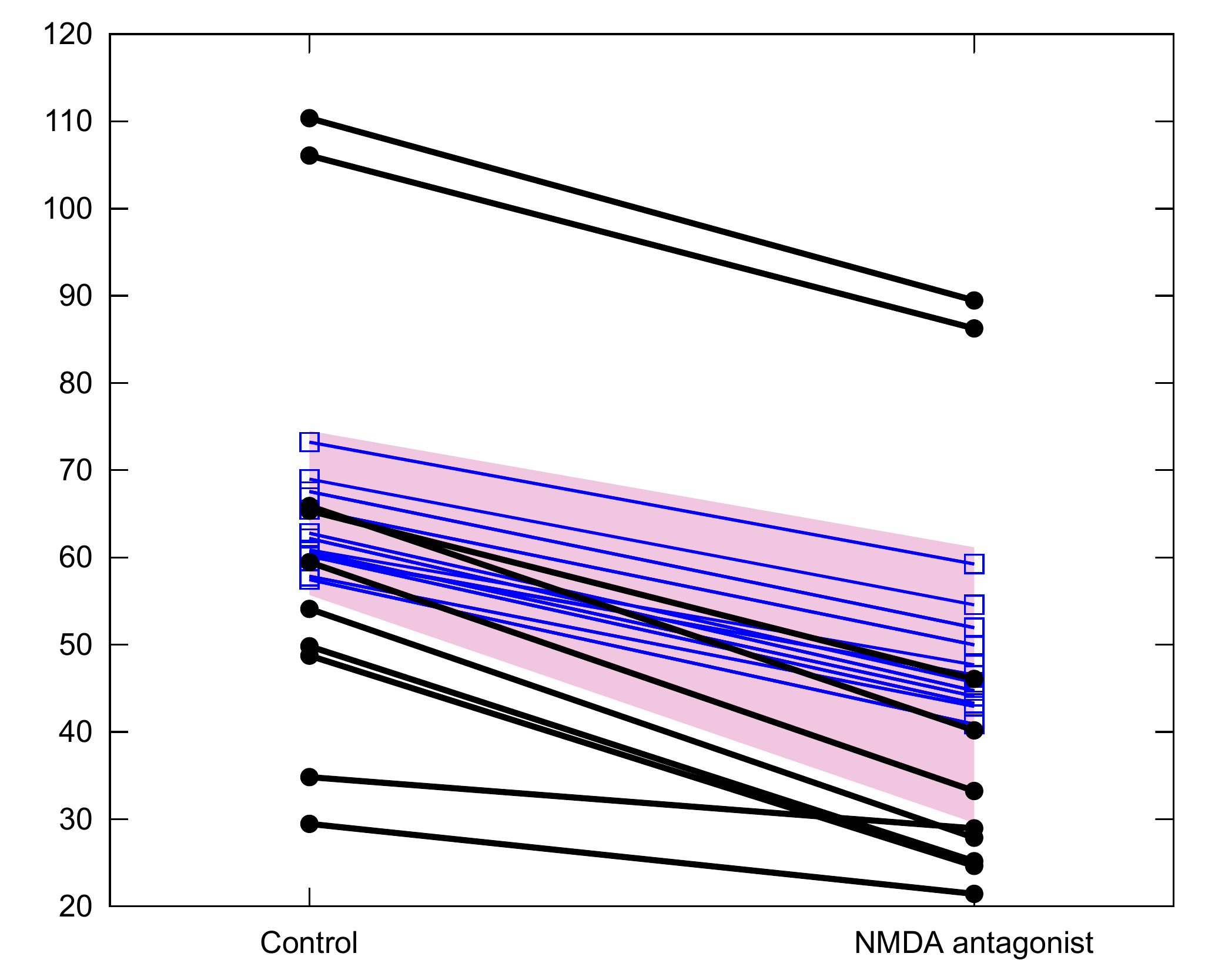}
\includegraphics[width=0.4\textwidth]{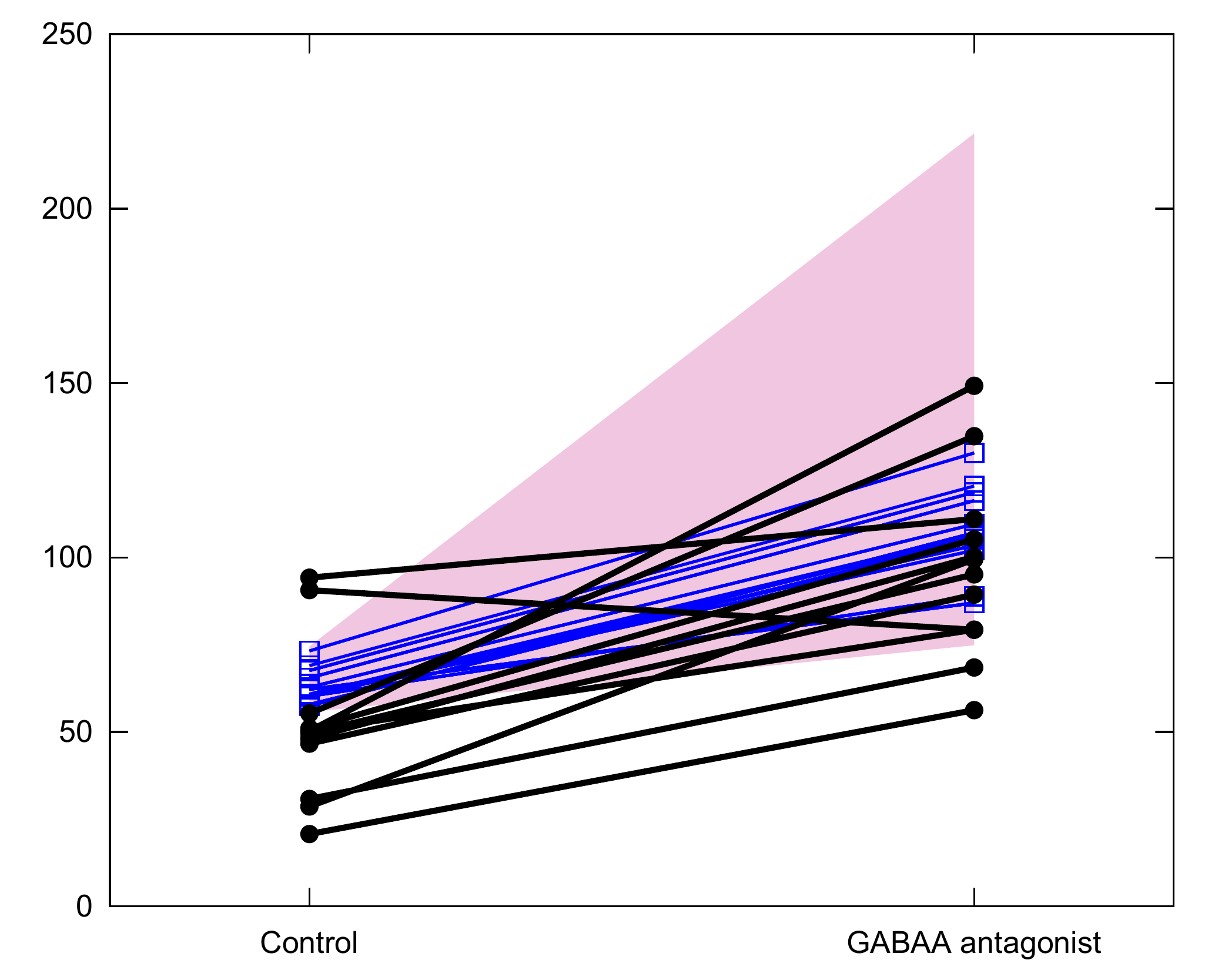} \\
\caption{GPe mean firing rates of each model (blue), plausible ranges allowed during optimization (pink) along the original single cell recordings (black) used to set them. The experimental data in black have been extracted from \cite{Kita04}.}
\label{fig:desact1}
\end{figure}

\begin{figure}[htpb]
\centering
E\includegraphics[width=0.4\textwidth]{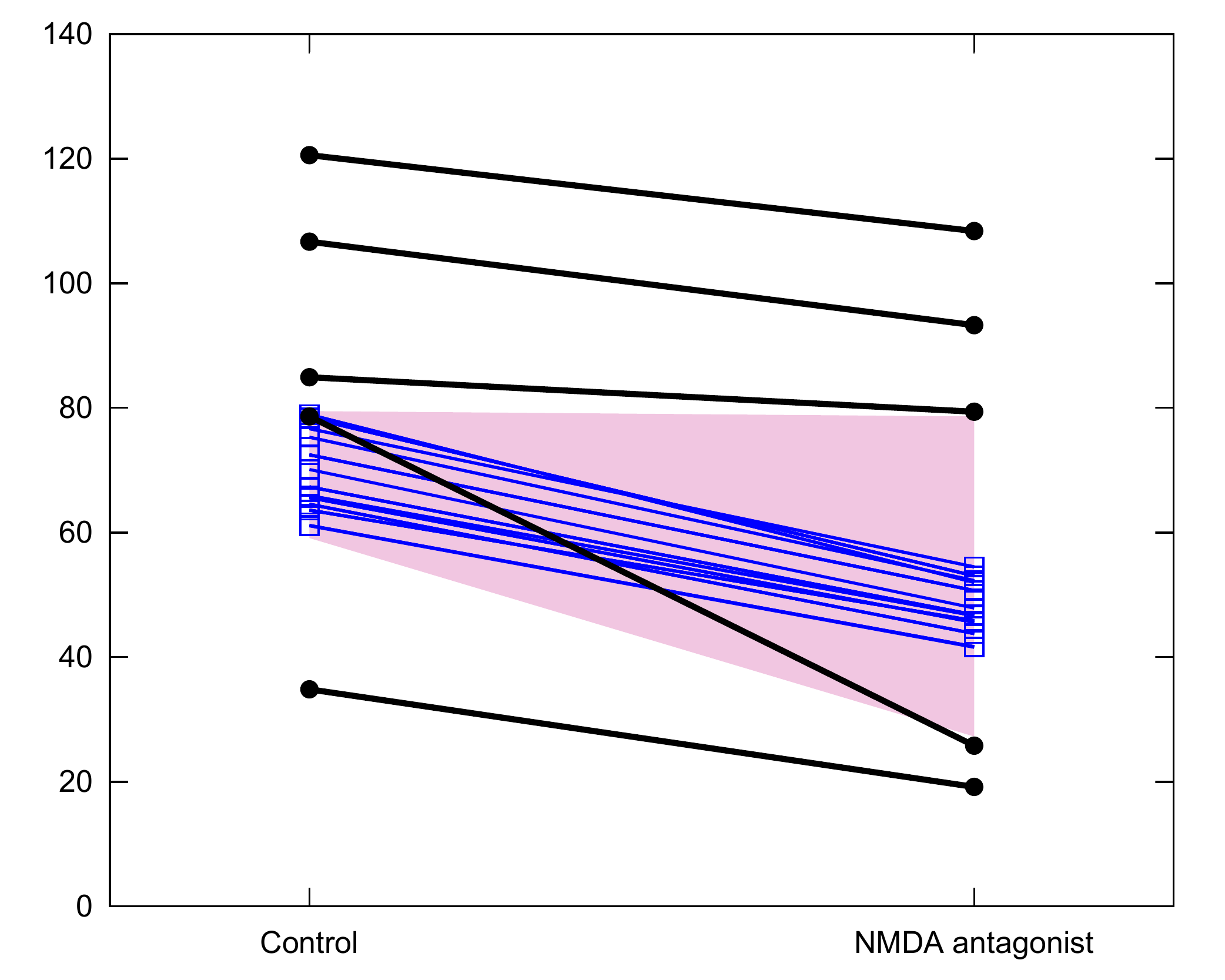}
F\includegraphics[width=0.4\textwidth]{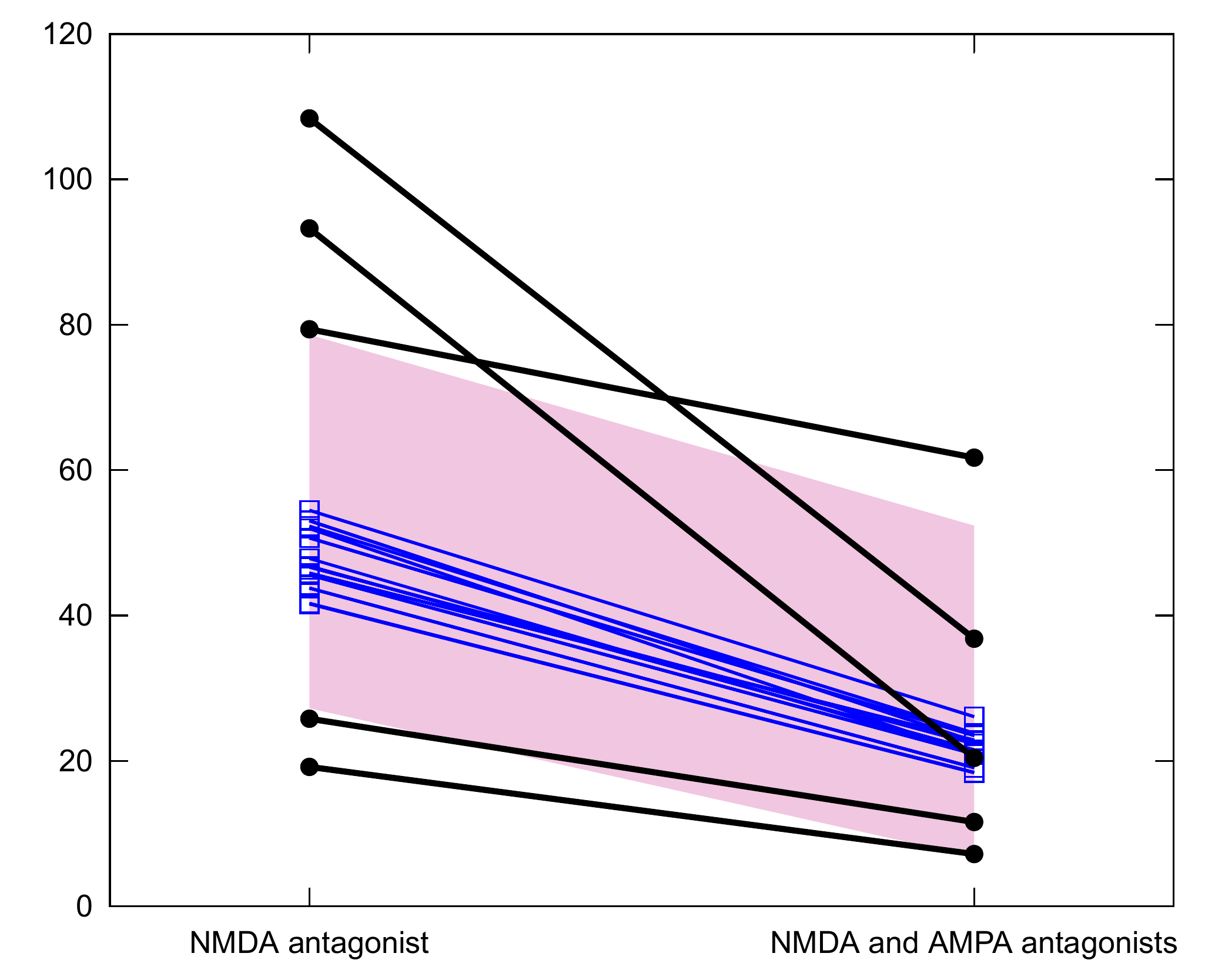}
G\includegraphics[width=0.4\textwidth]{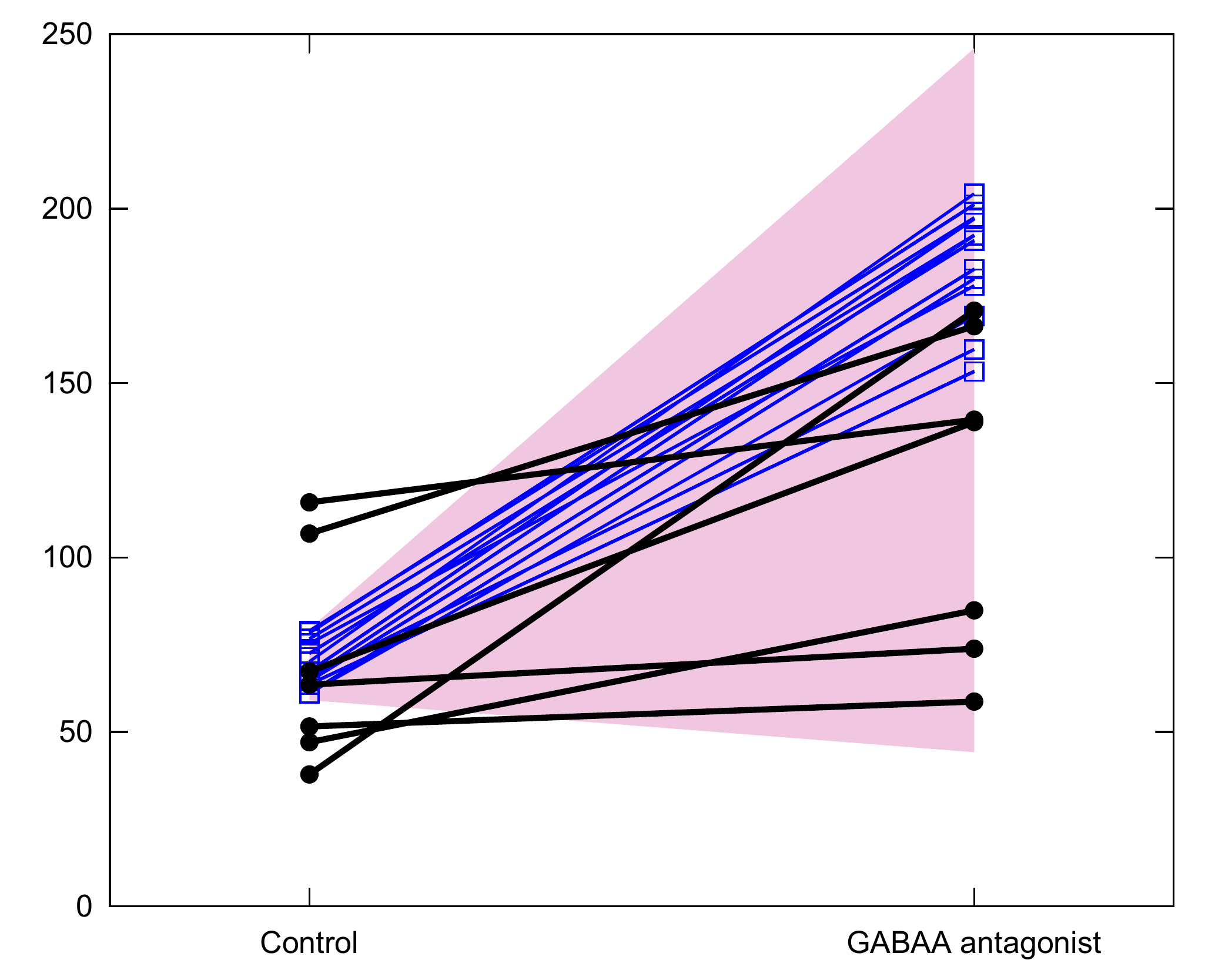}\\
\caption{GPi/SNr mean firing rates compared to experimental data, as in Figure \ref{fig:desact1}. The experimental data in black have been extracted from \cite{Tachibana08}.}
\label{fig:desact2}
\end{figure}

As stated above, the simulations of deactivations by antagonist experiments were maximized at 100\% for the electrophysiological score, which means they were perfectly reproduced according to the metric chosen. Moreover, the bulk of firing rates is never pressed against a plausible boundary (see Figure \ref{fig:desact1} for GPe and Figure \ref{fig:desact2} for GPi/SNr); if this was the case for one or more fit to a deactivation experiment, then this would have indicated that the optimization maximization of this part of the electrophysiological score conflicts with the rest of the electrophysiological and anatomical objectives.
This constitutes a strong validation of both these experiments and our models.


\notectable{
botcap,
caption = {Extreme values of the connection strength $f$ for the optimized models. These values include the relative number of neurons, different types of neurotransmitter and mean distance of receptors along the synapses. See methods for more details.},%
label = {tab:minmaxparams},notespar}{lrrr}{ 
\hl{Parameter} & \hl{Unit} & \hl{Min} & \hl{Max} \ML
Connection strength & ($\mu$V.s) & & \\  \cmidrule(r){1-2}
$f_{\mbox{\scriptsize CSN   }  \rightarrow \ \mbox{\scriptsize MSN }}$ &  &  1132.07     & 1465.67   \\
$f_{\mbox{\scriptsize CSN   }  \rightarrow \ \mbox{\scriptsize FSI }}$ &  &  777.39      & 1482.94   \\
$f_{\mbox{\scriptsize PTN   }  \rightarrow \ \mbox{\scriptsize STN }}$ &  &  1050.69     & 1116.78   \\
$f_{\mbox{\scriptsize MSN   }  \rightarrow \ \mbox{\scriptsize GPe }}$ &  &  12814.49    & 15504.70  \\
$f_{\mbox{\scriptsize MSN   }  \rightarrow \ \mbox{\scriptsize GPi }}$ &  &  14075.90    & 26790.20  \\
$f_{\mbox{\scriptsize STN   }  \rightarrow \ \mbox{\scriptsize GPe }}$ &  &  331.26      & 591.66    \\
$f_{\mbox{\scriptsize STN   }  \rightarrow \ \mbox{\scriptsize GPi }}$ &  &  180.80      & 289.79    \\
$f_{\mbox{\scriptsize STN   }  \rightarrow \ \mbox{\scriptsize MSN }}$ &  &  0.00        & 0.40      \\
$f_{\mbox{\scriptsize STN   }  \rightarrow \ \mbox{\scriptsize FSI }}$ &  &  0.00        & 11.36     \\
$f_{\mbox{\scriptsize GPe   }  \rightarrow \ \mbox{\scriptsize STN }}$ &  &  39.60       & 50.73     \\
$f_{\mbox{\scriptsize GPe   }  \rightarrow \ \mbox{\scriptsize GPi }}$ &  &  22.33       & 29.91     \\
$f_{\mbox{\scriptsize GPe   }  \rightarrow \ \mbox{\scriptsize MSN }}$ &  &  0.00        & 0.15      \\
$f_{\mbox{\scriptsize GPe   }  \rightarrow \ \mbox{\scriptsize FSI }}$ &  &  7.18        & 35.52     \\
$f_{\mbox{\scriptsize GPe   }  \rightarrow \ \mbox{\scriptsize GPe }}$ &  &  46.26       & 46.88     \\
$f_{\mbox{\scriptsize FSI   }  \rightarrow \ \mbox{\scriptsize MSN }}$ &  &  59.38       & 118.98    \\
$f_{\mbox{\scriptsize FSI   }  \rightarrow \ \mbox{\scriptsize FSI }}$ &  &  28.22       & 118.05    \\
$f_{\mbox{\scriptsize MSN   }  \rightarrow \ \mbox{\scriptsize MSN }}$ &  &  146.04      & 456.01    \\
$f_{\mbox{\scriptsize CMPf  }  \rightarrow \ \mbox{\scriptsize MSN }}$ &  &  29.55       & 93.16     \\
$f_{\mbox{\scriptsize CMPf  }  \rightarrow \ \mbox{\scriptsize FSI }}$ &  &  626.32      & 1449.21   \\
$f_{\mbox{\scriptsize CMPf  }  \rightarrow \ \mbox{\scriptsize STN }}$ &  &  316.55      & 424.82    \\
$f_{\mbox{\scriptsize CMPf  }  \rightarrow \ \mbox{\scriptsize GPe }}$ &  &  66.52       & 300.00    \\
$f_{\mbox{\scriptsize CMPf  }  \rightarrow \ \mbox{\scriptsize GPi }}$ &  &  177.51      & 299.95    \\
$f_{\mbox{\scriptsize PTN   }  \rightarrow \ \mbox{\scriptsize MSN }}$ &  &  20.07       & 37.56     \\
$f_{\mbox{\scriptsize PTN   }  \rightarrow \ \mbox{\scriptsize FSI }}$ &  &  14.97       & 163.48    \\
\LL}

%
%






\subsection{analysis of the free parameters}

\subsubsection{Connection strengths}

\begin{figure}[t]
\centering
\includegraphics[width=0.45\textwidth]{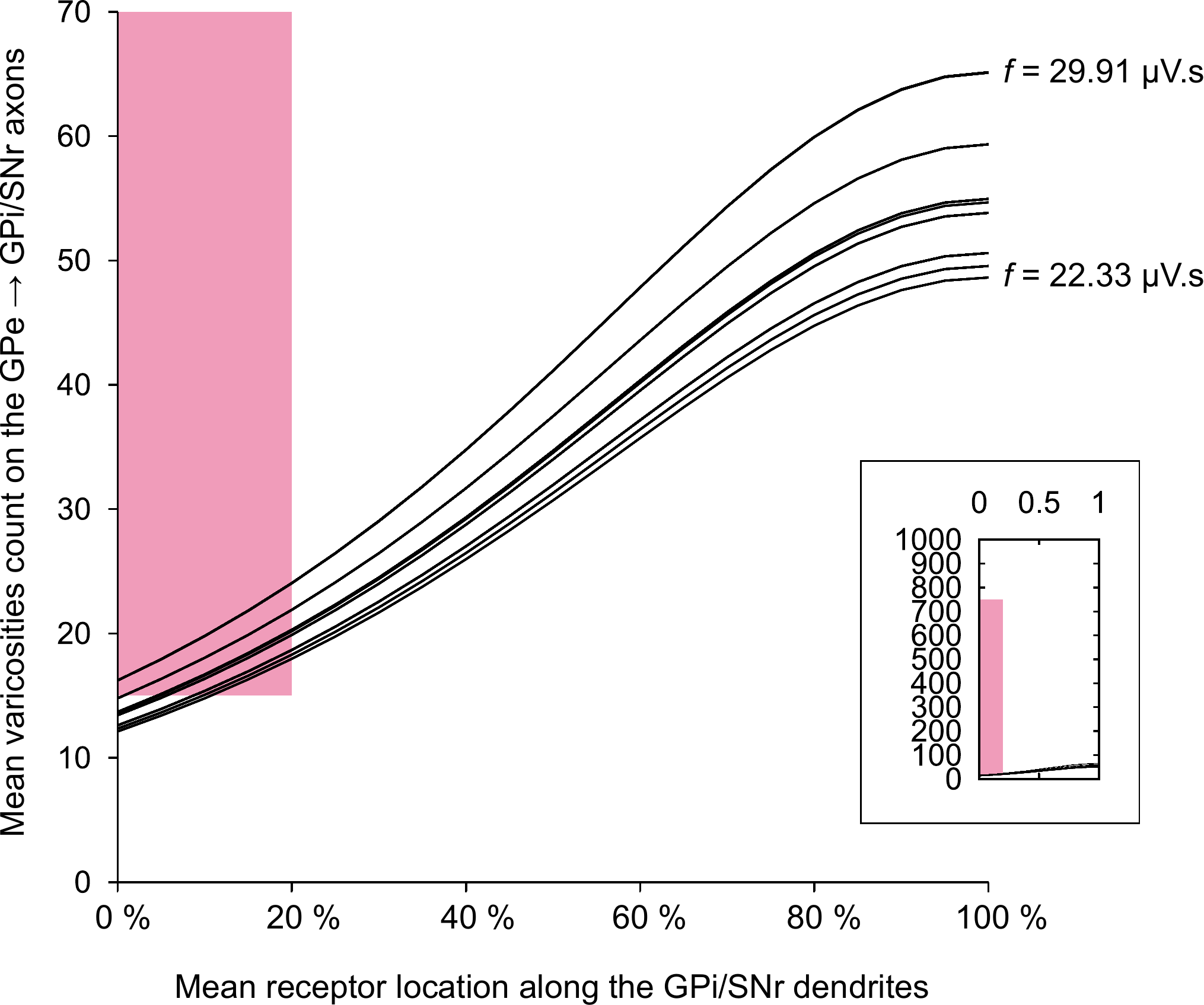}
\caption{Isoforces of the GPe~$\rightarrow$~GPi/SNr connection. The locations of the terminals along the dendrite are on the abscissa axis (100\% corresponds to terminal at the furthest location of the dendrites and 0 corresponds to terminals apposed on perikarya). The mean number of varicosities forming terminals are on the ordinate axis. The shaded area represents the plausible bouton counts and synaptic localization for a connection. One isoforce is calculated for each optimal solution, from its couple of parameters $(\alpha,p)$, and the values of extreme isoforces are indicated at the right. The inset represents the same isoforces at a different scale.}
\label{fig:isoforces}
\end{figure}

Of particular interest in the study of the BG is the GPe~$\rightarrow$~GPi/SNr connection, as the strength of this particular connection is very difficult to assess. Its strength has been explicitly questioned in \cite{Nambu08},
 and we hope to be able to bring here some elements of answer. We could indeed not constrain its mean bouton count per axon, resulting in the classification in a broad plausible range from 15 to 750. The connection isoforces at Figure \ref{fig:isoforces} shows that its bouton count is very low (i.e. 15 to 25) if GPe axons are targeting the soma of GPi/SNr neurons as is supported by \cite{Shink95} in primate. It could rise up to approximately 70 boutons if this connection was made at the extremity of the dendrites, but either way, the number of boutons would be categorized in the low range (see methods). However, the strength of this connection is overall of the same order of magnitude that of the GPe~$\rightarrow$~STN connection (c.f. Table \ref{tab:minmaxparams}). A finer analysis, solution by solution, of the ratio $f_{GPe \rightarrow GPi/SNr}$~/~$ f_{GPe \rightarrow STN}$ shows that it ranges from 46\% to 75\%. 

The general consensus in the understanding of the basal ganglia have been to consider the direct/indirect pathways as mirror pathways, targeting with the same strength the GPe and GPi/SNr. In the results presented at Table \ref{tab:minmaxparams}, the ranges of MSN~$\rightarrow$~GPi/SNr is higher than the ones of MSN~$\rightarrow$~GPe. Moreover, the $f_{MSN \rightarrow GPi}$~/~$f_{MSN \rightarrow GPe} $ ratios range from 110\% to 200\%. Together with previous results from a tracing study revealing that striatal projection neurons display more varicosities in the GPi/SNr than in the GPe \citep{Levesque05b}, this finding calls for a reappraisal of the relative direct/indirect pathways strength, with a direct pathway that could be twice as effective as the indirect pathway.
In line with the previous result, the STN~$\rightarrow$~GPe and STN~$\rightarrow$~GPi/SNr connections are not of similar strengths, the STN~$\rightarrow$~GPe connection being more potent than the STN~$\rightarrow$~GPi/SNr connection. The ratios $f_{STN \rightarrow GPe}$~/~$f_{STN\rightarrow GPi/SNr}$ are between 182\% and 270\%. This is at odds with with the accepted view that these projections are parallel, hence similar in strength.

The GPe recurrent projection (GPe~$\rightarrow$~GPe) is set to a rather precise strength. This connection is under a lot of pressures during the optimization process, as it has a direct impact on the firing rate of the GPe, which is at the center of the BG. Whether this reflects a difficulty to explore the parameter space for this connection or whether its value is heavily constrained by the antagonist studies used in the optimization, four of which involve injection and recording in its vicinity, is unknown at this point. We discuss in particular this connection later on.

GPe and STN connections to MSN are very weak, as the corresponding values were the lowest of the whole connection parameters. This is explained in part by the low synaptic counts found during the optimization, but also because the connection strength takes into account the relative number of neurons, which are more numerous by several order of magnitudes for the MSN. On the other hands, GPe and STN connection to FSI were potent and in a position to influence the whole circuitry of the BG.

The isoforces of these connections are not discussed any further here, and are available as supplementary materials in Appendix (Figure \ref{fig:allisoforces}).

\subsubsection{Striatal gabaergic interneurons firing rates parameters}

Due to the lack of data for the primate \citep{Berke12}, the firing rates of the FSI was not tightly constrained.
Optimal models have firing rates from 4 to 18 Hz, which occupy most of the plausible bounds of [0, 20 Hz]. However, their maximal firing rate, which was constrained by the construct objective into the wide [200, 500 Hz] range, has values only in the [200, 253 Hz] range, predicting that in average the FSI maximal firing rate is quite low in primate compared to other species.

\subsubsection{$\theta$ parameters}

The $\theta$ parameters correspond to the neurons excitability and were set freely during the optimization, because the only available experimental data concerns {\it in vitro} rodent slice, and because it does not permit to constrain this parameter unambiguously. The obtained results are shown in Table \ref{tab:thetas}, and are discussed in the light of previous models in section \ref{sec:fixparams}.

\notectable{
botcap,
caption = {Minimum and maximum of the difference between resting and firing potentials from our study, and their values from two previous mean field models.},%
label = {tab:thetas},notespar}{lrrrrr}{ 
  \hl{Parameter  (mV)}     &    \hl{Min}  & \hl{Max}    &  \hl{from} \citable{VanAlbada09a} &  \hl{from} \citable{Tsirogiannis10}  \ML
  $ \theta_{\mbox{MSN}} $     &   28.5     & 30        &    19  & 27   \\
  $ \theta_{\mbox{FSI}} $     &   11.7     & 20.2      &    N/A & N/A  \\
  $ \theta_{\mbox{STN}} $     &   24.6     & 26.1      &    10  & 18.5 \\
  $ \theta_{\mbox{GPe}} $     &   6.9      & 11        &    9   & 14   \\
  $ \theta_{\mbox{GPi}} $     &   5        & 6.3       &    10  & 12
\ML}


\subsection{Selection capability}

\begin{figure}[htb]
\centering
Ctx\includegraphics[width=0.35\textwidth]{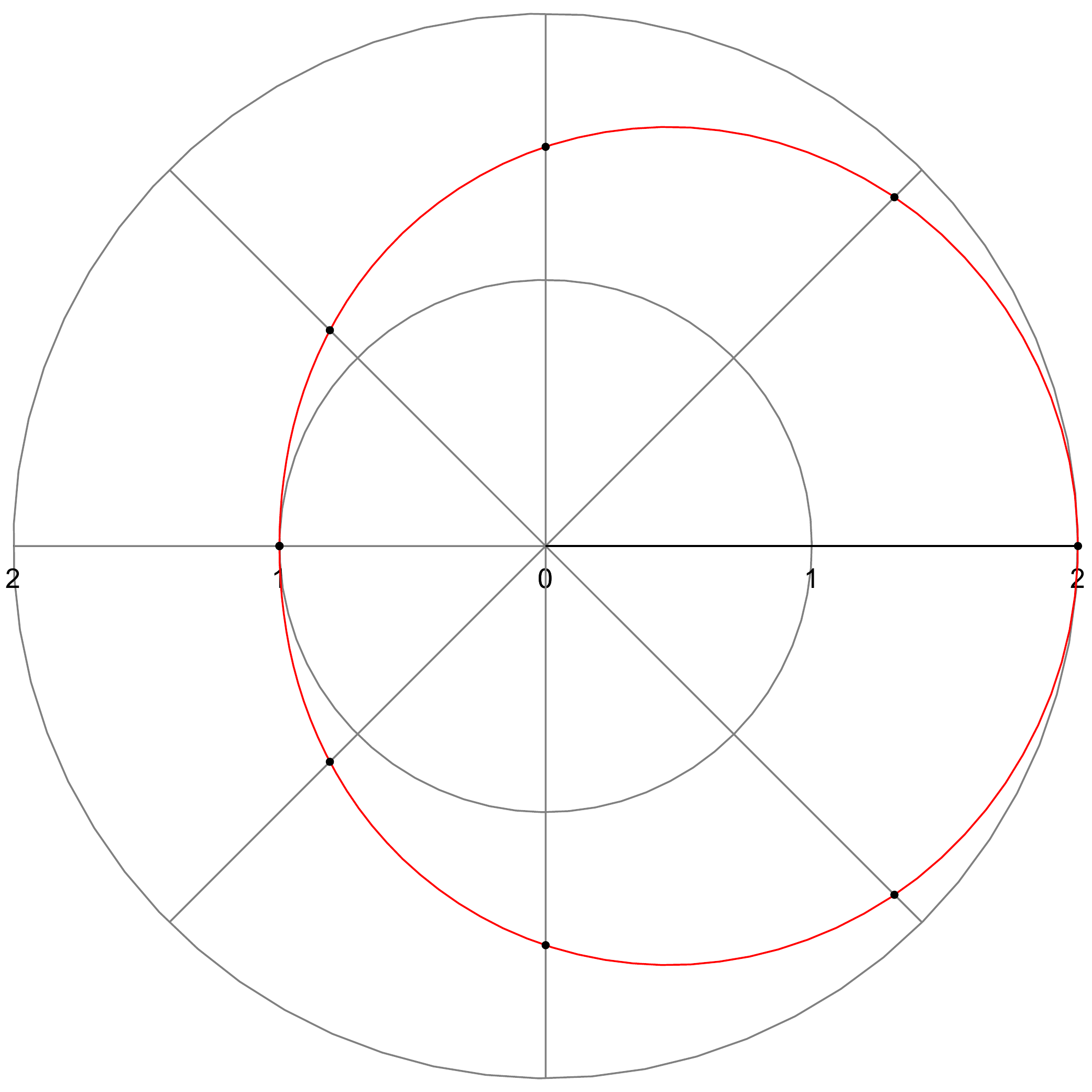}
MSN\includegraphics[width=0.35\textwidth]{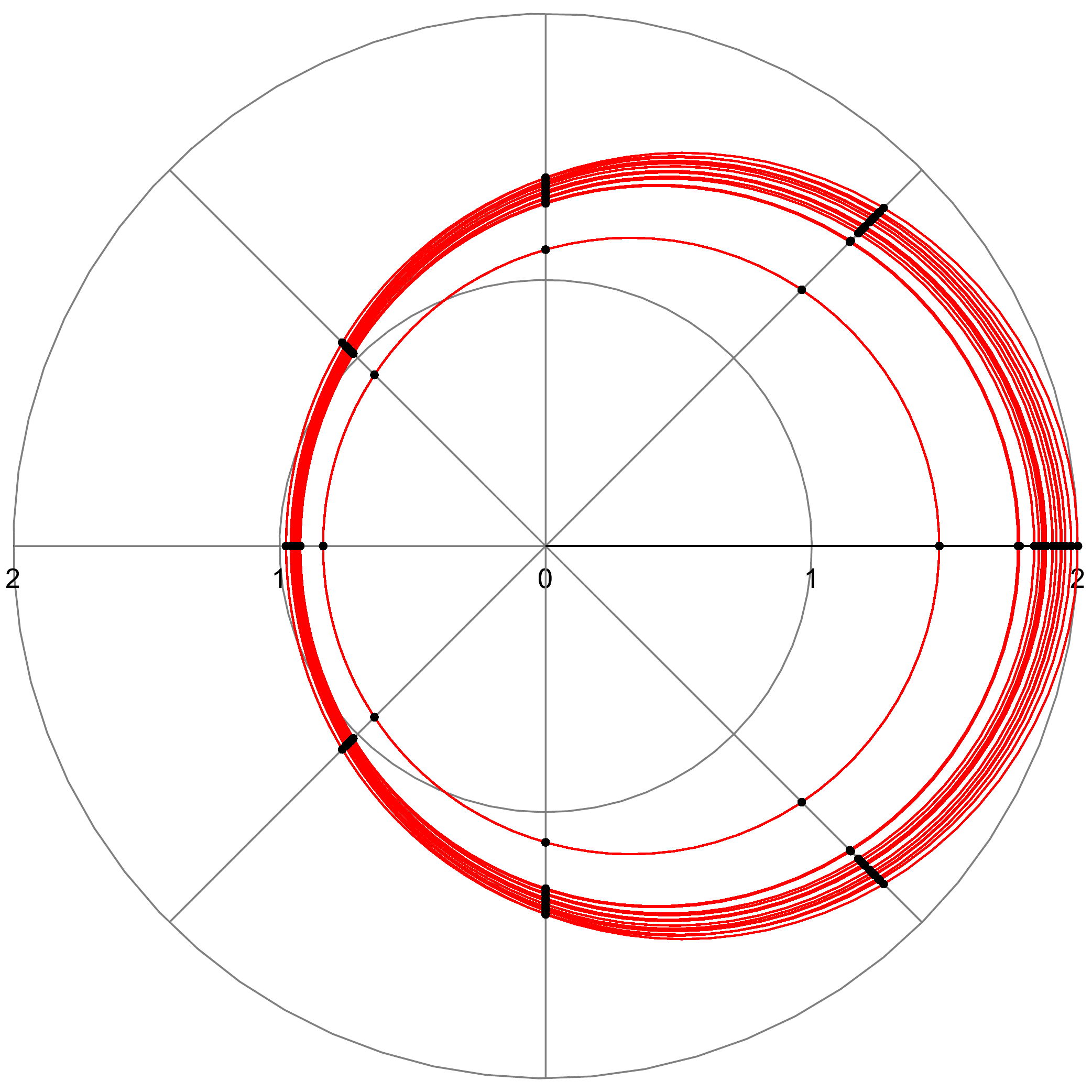}
FSI\includegraphics[width=0.35\textwidth]{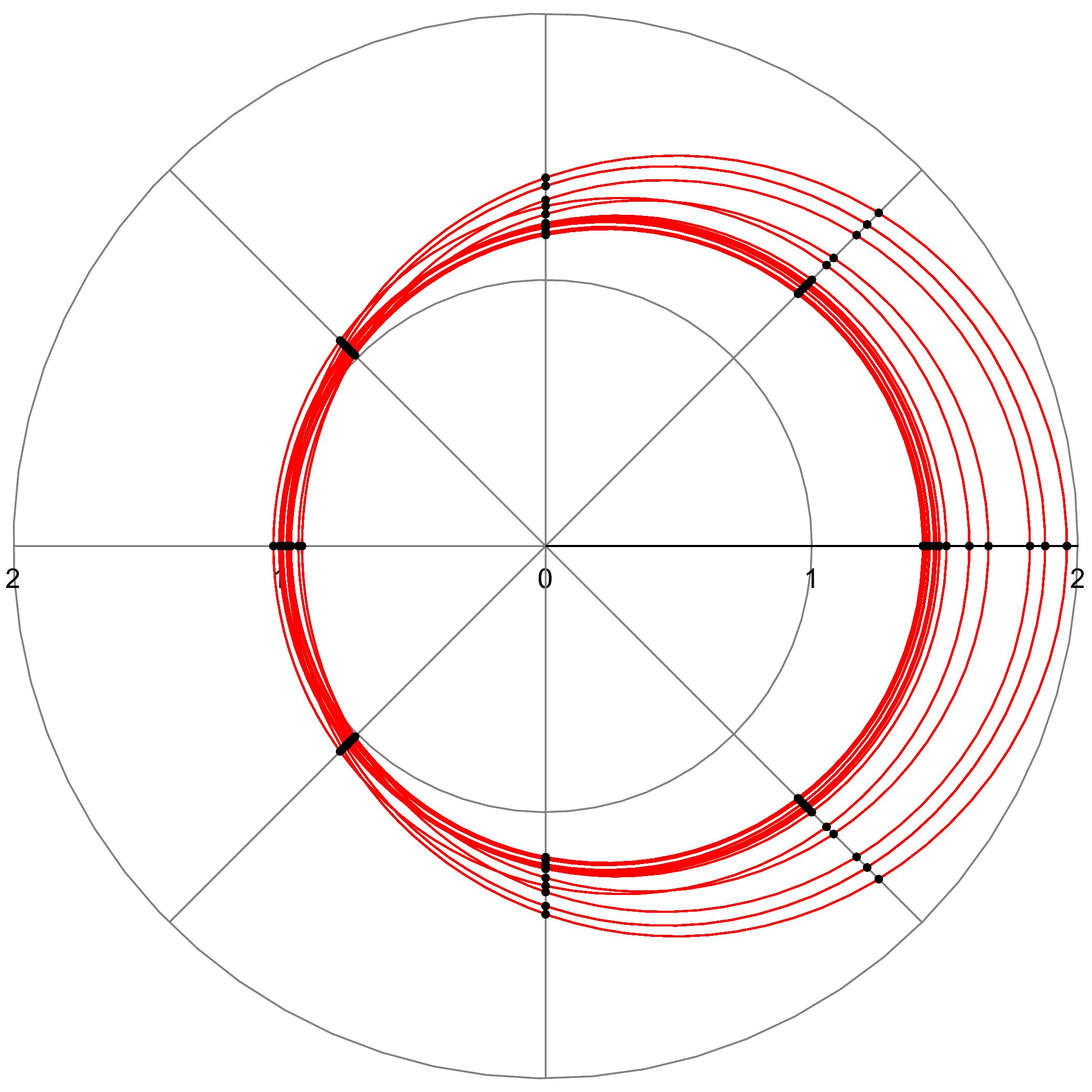}\\
\caption{Mean firing rates of the models during selection according to the direction, normalized relatively to the activity at rest (corresponding to the dark inner circle). The preferred direction at 0 degree is along the abscissa axis, to the right.}
\label{fig:dir1}
\end{figure}

\begin{figure}[htb]
\centering
STN\includegraphics[width=0.35\textwidth]{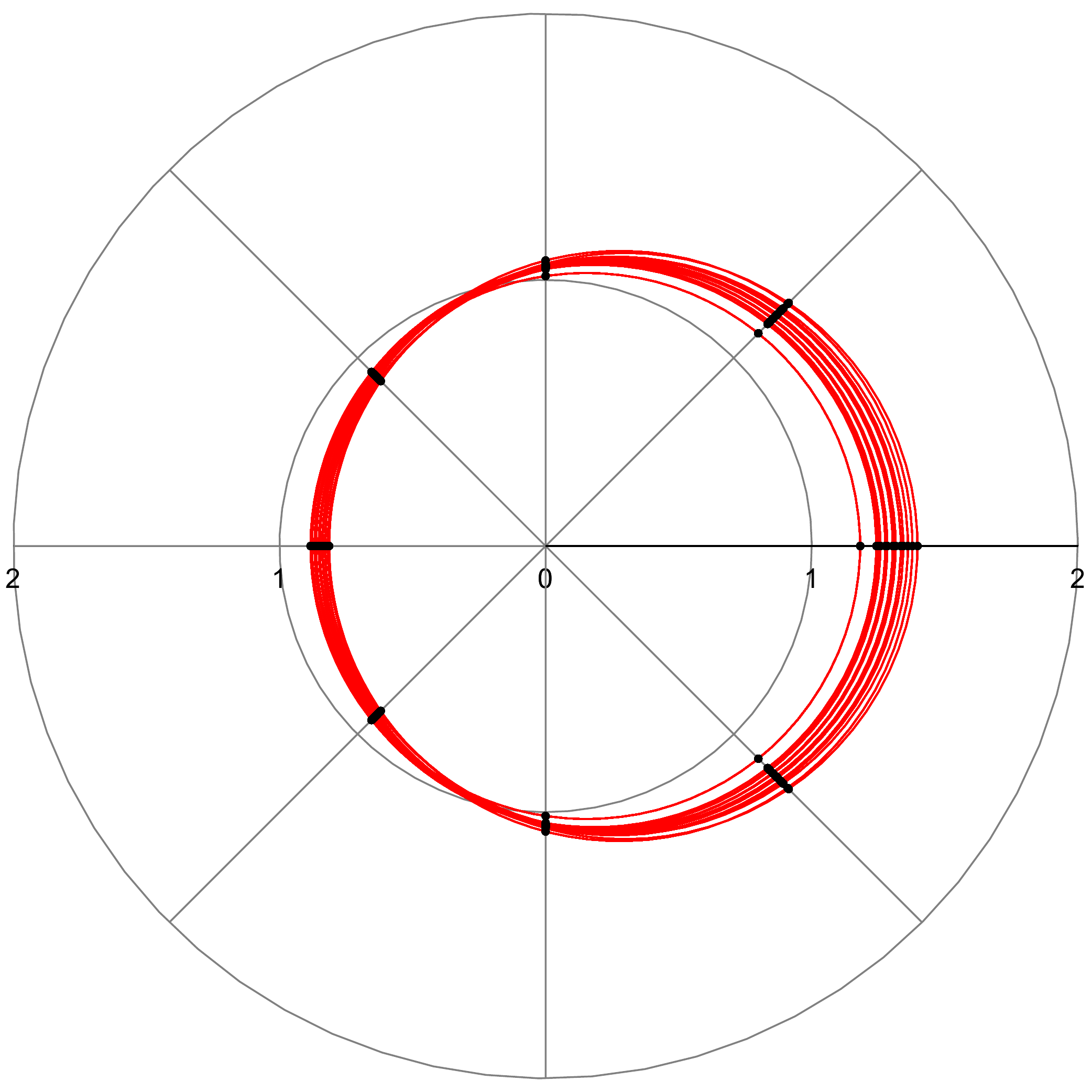}
GPe\includegraphics[width=0.35\textwidth]{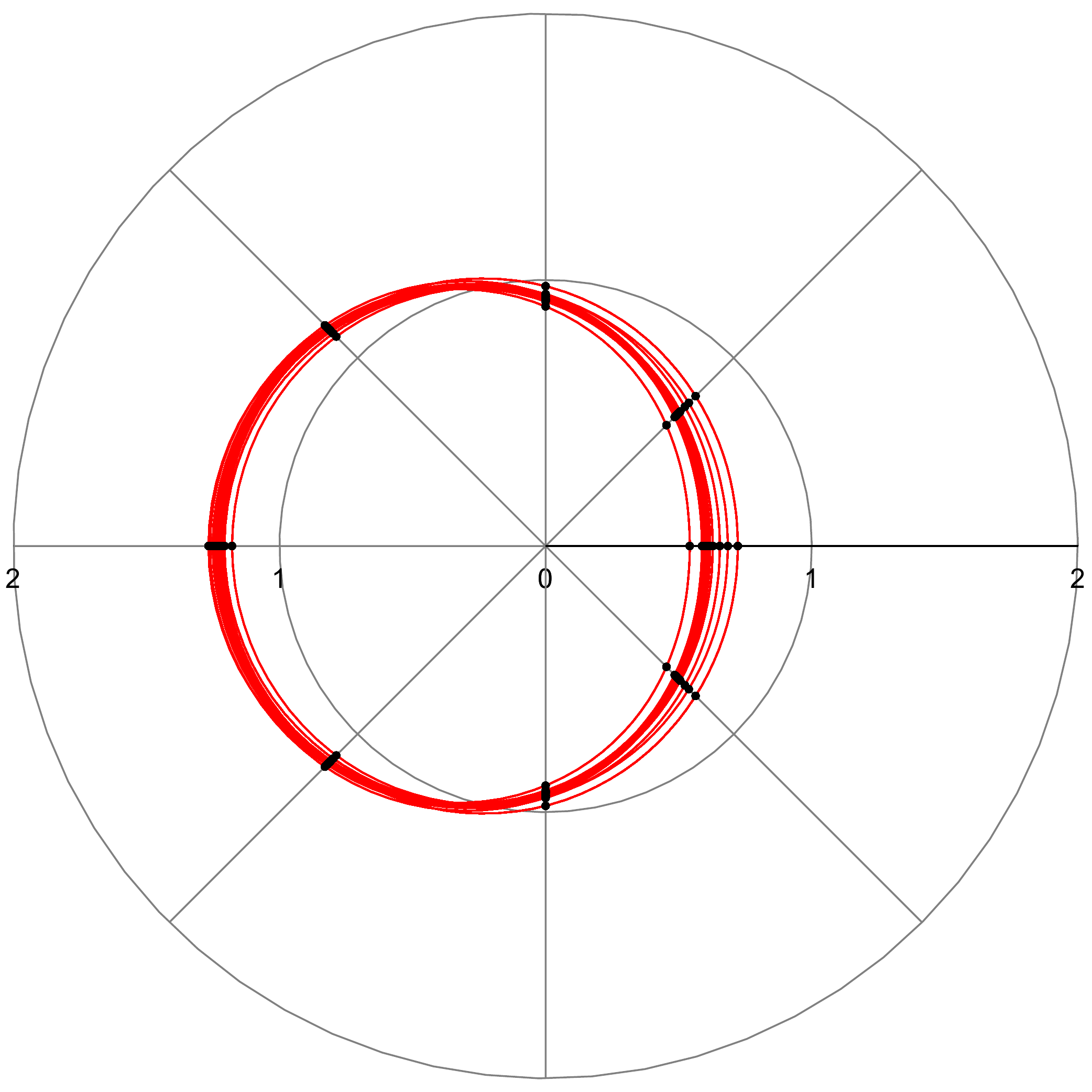}
GPi/SNr\includegraphics[width=0.35\textwidth]{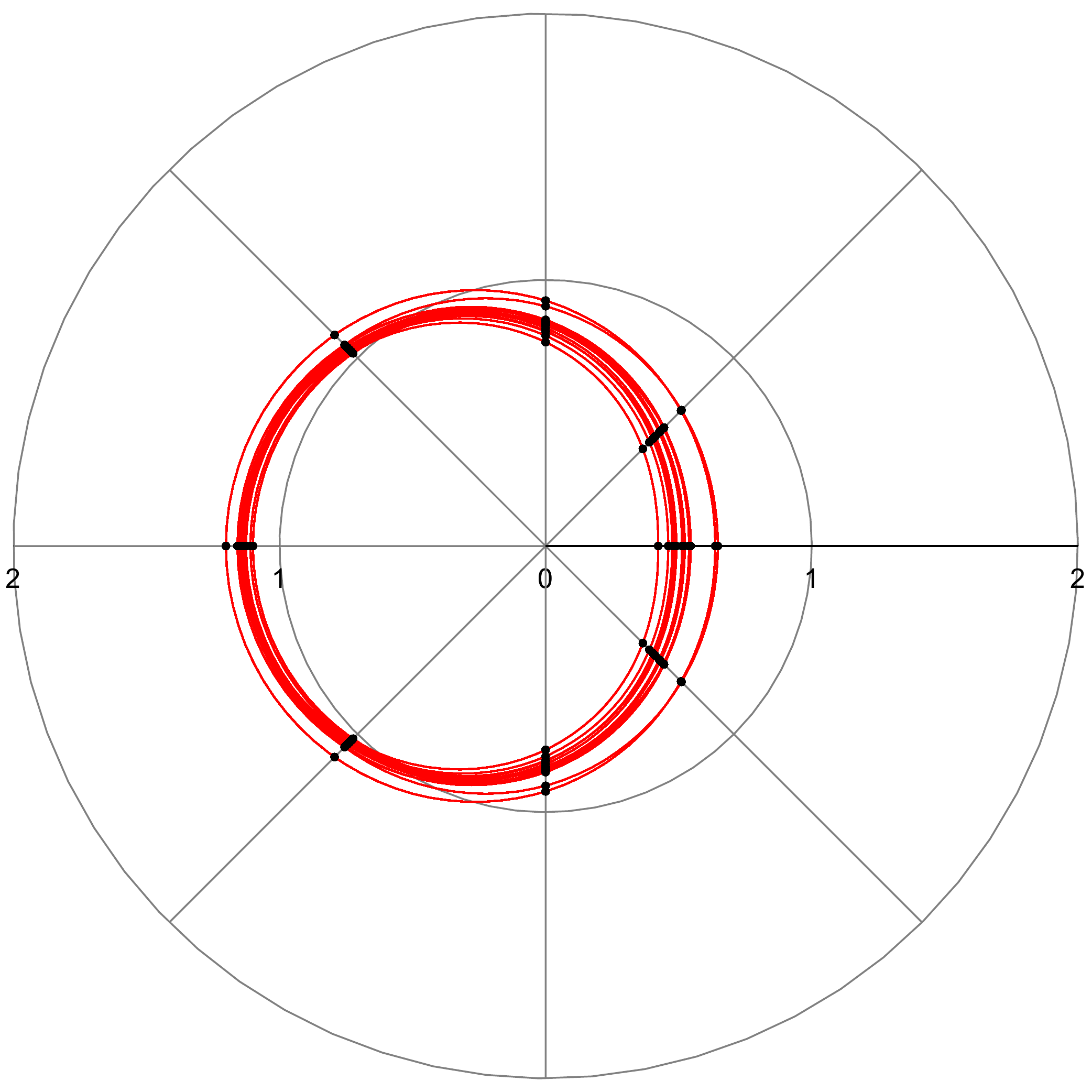}\\
\caption{Mean firing rates of the models during selection according to the direction (see Figure \ref{fig:dir1}). In these traces, the GPe~$\rightarrow$~GPi/SNr connection was in the form 1-all.}
\label{fig:dir2}
\end{figure}

We submitted the optimal models to the directional selection test as detailed in methods. The firing rates converged to a stable output in less than 50 ms. Although the inputs were sent only from the cortico-striatal pathway, the directionality could be observed in every nucleus, including the STN. Moreover, the neuron cluster corresponding to the best directions in GPi/SNr have lower than baseline activity while the other clusters have near-baseline or higher activities, which implements a selection by inhibition (Figures \ref{fig:dir1} and \ref{fig:dir2}).

\todothese{Faire des distances de shapley pour determiner quels sont les nuclei qui sont impliques dans la selection}


All the connection schemes (either diffuse or focused) could be constrained apart from GPe~$\rightarrow$~GPi/SNr, as the only single axon tracing study available for the primate does not allow to decide between these two patterns \citep{Sato00a}. We tend to think that this connection operates on a diffuse manner because the GPe appears to have activities mirroring the GPi/SNr, and thus GPe~$\rightarrow$~GPi/SNr focused inhibitions would cancel selection performed at the GPi/SNr level. Yet we can not rule out the possibility that this connection is focused because of the lack of conclusive anatomical evidence.

\begin{figure}[htb]
\centering
\includegraphics[width=0.45\textwidth]{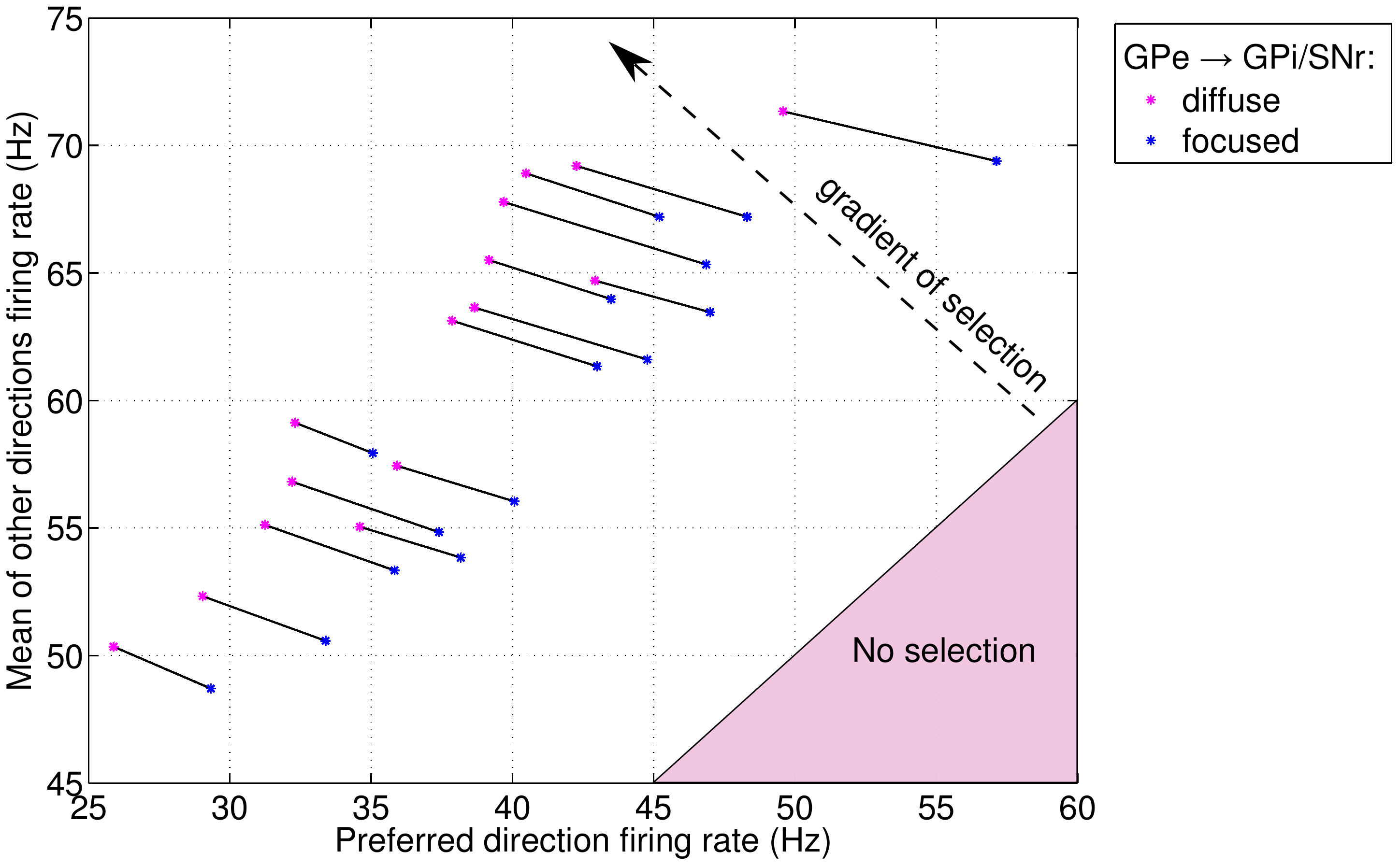}
\caption{For each obtained solution, firing rate for the preferred direction (on the abscissa axis) versus the opposite direction (on the ordinate axis) of BG output. Two cases for the GPe~$\rightarrow$~GPi/SNr connection pattern are represented: diffuse (red dots) and focused (blue dots). Lines linking two points indicate that they were calculated with the same set of parameters. The area corresponding to the absence of selection is shaded, and the gradient of better selection in the other part of the graph is represented by an arrow.}
\label{fig:quantif}
\end{figure}

To quantify the effect of the GPe~$\rightarrow$~GPi projection pattern on the quality of the selection, we looked at the firing rate in output of the GPi/SNr for the preferred direction versus the mean firing rates of the other directions, for the diffuse and focused GPe~$\rightarrow$~GPi/SNr connectivity patterns (Figure \ref{fig:quantif}). This shows that the pattern has an influence, and a focused pattern improves selection on both indicators: the preferred direction has a lower firing rate while the mean of the other directions have a larger firing rate.

\begin{figure}[htb]
\centering
\includegraphics[width=0.35\textwidth]{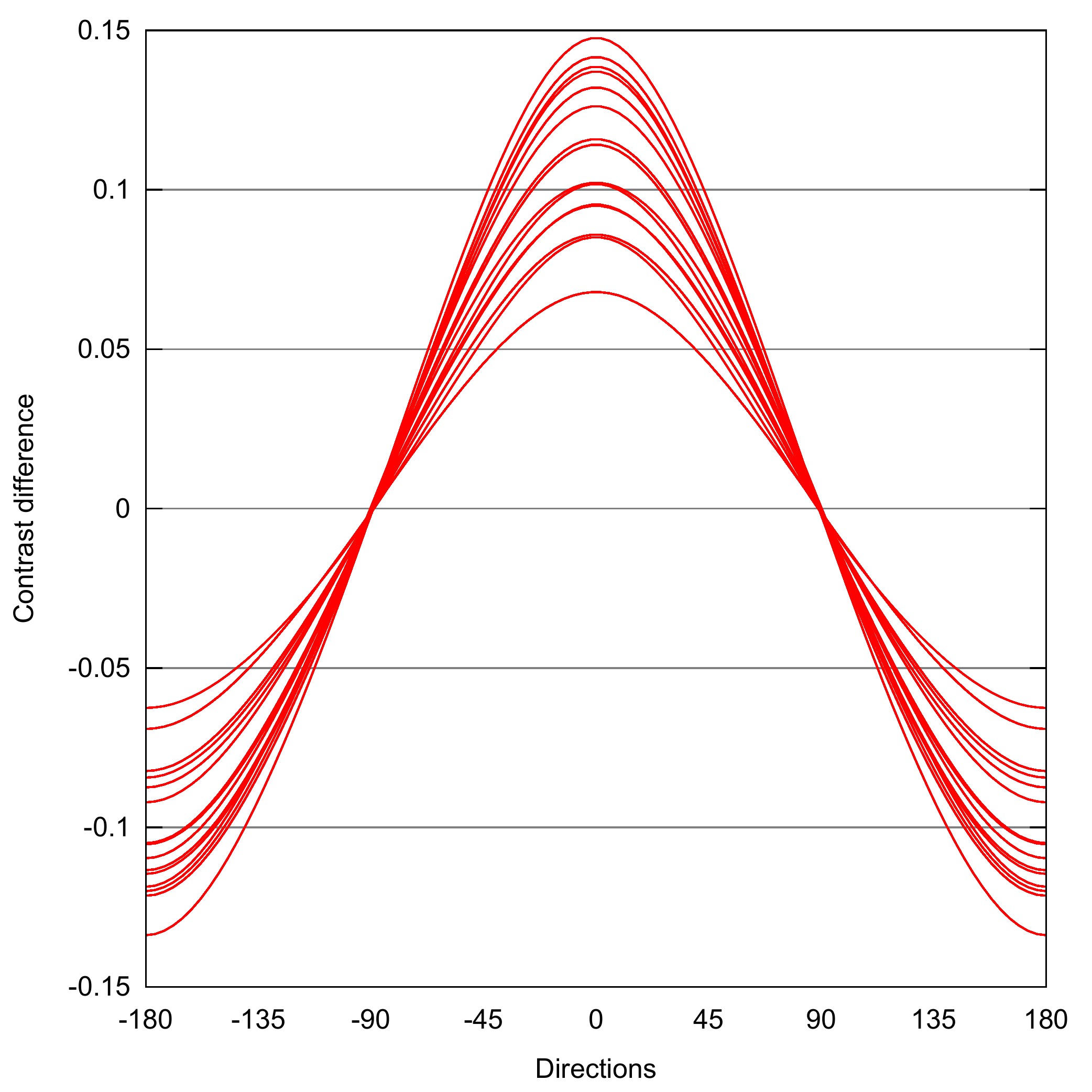}
\caption{For all optimized models, plot of difference of contrast values $c_i$ between diffuse and focused connectivity for the GPe~$\rightarrow$~GPi/SNr connection for all directions. Positive values denote better selections in the diffuse case compared to the focused case, whereas negative values denote better deselections. This graph shows that for all models, a diffuse GPe~$\rightarrow$~GPi/SNr projection leads both to a better selection of directions close to the preferred directions, and a better deselection of directions close to the opposite directions. See methods for the mathematical formulation of the indicator.}
\label{fig:contrasts}
\end{figure}

Furthermore, the values of the direction-dependent contrast differences $c_i^{\mbox{\tiny \ \scriptsize (diffuse)}} - c_i^{\mbox{\tiny \ \scriptsize (focused)}}$ indicate that the influence of the pattern is direction-dependent. Indeed, the difference of the contrast between the two patterns shows that the selection of the directions corresponding to the preferred choice is enhanced while the opposite directions are more deselected (see Figure \ref{fig:contrasts}).


\section{Discussion}

We studied the physiology and function of the BG through the design of models respecting as much as possible anatomical data (the {\it construct} objective) and electrophysiological data (the {\it face} objective). We focused in particular on the GPe and GPi/SNr, as their parameters were the most constrained by the electrophysiological data fitted \citep{Kita04,Tachibana08}. The main outcomes of our study are as follow:

\begin{itemize}
\item[$\bullet$] Available anatomical and electrophysiological data on the GPe~$\rightarrow$~GPi/SNr connection are consistent. The strength of this connection does not support opposed activities in the GPe and GPi/SNr, and a diffuse projection pattern is more efficient for the overall selection operated in BG.
\item[$\bullet$] Symmetry is not the rule in the strengths of the BG connections. In particular, both STN and MSN target more the GPe than the GPi/SNr. 
\item[$\bullet$] Selection arises from the structure of the BG, without properly segregated direct and indirect pathways and without specific inputs from pyramidal tract neurons of the cortex.
\end{itemize}

\subsection{GPe and GPi/SNr relations}

The strength of the GPe~$\rightarrow$~GPi/SNr connection was not precisely constrained for the optimization, due to the lack of precise anatomical data on it. The GPe targets the GPi/SNr proximally \citep{Shink95}, which would suggest a potent connection. However, they both have comparable high firing rates at rest (c.f. Table \ref{tab:firing_rates}), and they have
correlated increase/decrease at rest \citep{Elias08}, in directional arm reaching tasks \citep{Georgopoulos83a,Turner97} and in eye closures \citep{Adler10}. These electrophysiological data
would on the contrary indicate that this connection is not potent, and this apparent paradox has been puzzling scientists for many years \citep{Mink96,Nambu08}. By construction, the optimized models presented
here are compatible with both the firing rates at rest and the known anatomical data, thus showing these are consistent despite their apparent contradiction.


Our study first provides a quantitative prediction on the strength of the GPe~$\rightarrow$~GPi/SNr connection as well as all the others connection (summarized in Table \ref{tab:minmaxparams}):
the strength of this connection can either be considered as weak compared to the striatal input to the GPi/SNr, or as quite potent in comparison to other GPe efferences (e.g. it is only roughly half as potent as the GPe~$\rightarrow$~STN connection).
We hope that these quantitative indications will make further interpretation of this connection easier.

We also characterized the influence of this connection on the selection capability. GPe and GPi/SNr exhibit mirroring activities in a selection task (see Figures \ref{fig:dir1} and \ref{fig:dir2}), which is a prediction at odds with other computational models assuming opposed activities (e.g. \cite{Rubin04,Shouno09}).
Furthermore, in our study the focused pattern for this connection has a negative impact on the selection capability compared to the diffuse pattern, bringing support to a previous work postulating that this connection is diffuse \citep{Girard08}. As this connection has an overall negative impact on selection (data not shown) one hypothetical purpose for it could be to prevent the GPi/SNr to have a firing rate too high in order to save energy. In regard to this hypothesis, the GPe~$\leftrightarrow$~STN interconnection could be mainly serving a purpose not directly linked to selection \citep[e.g. it has been proposed that it acts as a random generator for exploration in][]{Magdoom11}, and as a side benefit the GPe~$\rightarrow$~GPi/SNr connection could keep the GPi firing rate at reasonable values.


\subsection{Anatomy}

We were able to provide a novel account of the relative internuclei connection strength in the whole BG, and not only for the GPe and GPi/SNr interactions.

%
%
%

We suggest that symmetry is not the rule for the BG. Indeed, the MSN should be targeting more the GPi/SNr than the GPe, with a ratio in the 110\% - 200\% interval. This is in line with an axonal tracing study which show that more axonal boutons from MSN target the GPi/SNr than the GPe \citep{Levesque05a}. Moreover, the STN should be targeting more the GPe than the GPi/SNr, with plausible ratios ranging from 182\% to 270\%. It has been assumed since the discovery that the STN targets reciprocal regions of the GPe and GPi/SNr \citep{Shink96}, that its projections toward both compartment of the globus pallidus are symmetrical in strength. Our results call into question this simplification, and challenge an interpretation that has been prevailing in computational models for 20 years.

The results obtained on the striatal afferent from the other BG nuclei are probably less precise because less constrained by the antagonist deactivation studies used, which concerned both GPe and GPi/SNr. Nonetheless, due to the difference of neuron numbers that was taken into account in the models, the connection from GPe and STN toward MSN was found to be negligible, while GPe~$\rightarrow$~FSI was found to be potent and STN~$\rightarrow$~FSI could either be potent or negligible. The connection from the STN hence needs more studies to determine which neuronal type it targets in the striatum, as it would likely have unexplored functionnal implications if the FSI are specifically targeted. In addition, the connections from GPe and STN could {\it also} be targeting neurons of the striatal patch that we excluded from our models, and as they are less numerous - approximately 10\% of the total striatal count, see \cite{Johnston90} - they could thus have a greater functional impact.

The GPe~$\rightarrow$~GPe has been set at very precise values, but we cannot rule out the possibility that it results from the high number of constraints that we put on it, or from a lack of exploration around the solutions by the optimization algorithm.
The interpretation of this result is unfortunately hard to make, because the GPe can not be considered as a monolithic nucleus but it should rather be considered as a complex assembly of several different neuronal types \citep{Cooper00}, for which rodents study have revealed that they do not share an equal number of axon collaterals. Indeed, both the dendritic arborization and number of axon collaterals depend on the neuron distance from the GPe medial border \citep{Sadek07}. Furthermore, the axons of neurons targeting solely the striatum have been found to exhibit more recurrent collaterals than the others \citep{Mallet12}.
As we modeled the GPe with the simplest choice of considering its neurons as a uniform population here, the particular organization of the GPe micro-circuitry is out of reach. Exploring the possible subdivisions of this highly complex nucleus could be the subject of future work.

\subsection{Selection}

After including as much data as possible in a model of the BG and optimizing it to the best extent, we obtained solutions that have consistent parameterizations, and which were found to perform selection. In other words, the whole anatomically plausible set of parameters obtained results in a mean-field model doing selection if this model permits to recreate the electrophysiologically plausible activity. This is different from previous results, which showed that
(a)
depending on the BG anatomical details included and on the specific parameterization chosen, different selection scheme could be performed \citep{Berns94,Gurney01a,Girard08};
or (b)
with a specific fine-tuned set of parameters fitting electrophysiological studies, models also do selection \citep{Humphries06};
or again (c)
 it is possible to map an optimal decision making algorithm on a simplified structure of the BG \citep{Bogacz07}.
 While the idea that the BG perform selection is not new \citep{Mink96,Redgrave99}, our results bring an original contribution to it, by showing that respecting a lot of anatomical and electrophysiological data is {\it sufficient} for the emergence of the selection ability. This result can thus be viewed as a validation of the theory that selection occurs in the BG.


 Since the very first box-and-arrows models of the BG \citep[e.g.][]{Albin89}, the subdivision in direct and indirect pathways has taken a central place in the understanding of their function in the normal state and in pathophysiology. However, as we argue in the methods for the primates, the correspondence between D1 and D2 receptors and the direct and indirect pathway is unclear, and moreover the segregation into a direct and an indirect pathway is not supported by single axon tracing studies. Our results here show that their explicit segregation is not necessary for selection. This is a first step towards future works exploring the mixed influence of dopamine on MSN when D1 and D2 receptors are partially overlapping and in absence of real segregation in their pallidal targets, in conformity with what is known in the primate.

Furthermore, in the primate, directionality in a two dimensional reaching task has been found in most populations of the BG: in MSN \citep{Hori09}, in STN \citep{Georgopoulos83a}, in GPe and GPi \citep{Georgopoulos83a,Turner97}. This is in line with the activity of the optimized models (c.f. Figures \ref{fig:dir1} and \ref{fig:dir2}). 
%
Interestingly, the directionality that we observed in STN does not result from a specific signal conveyed by the cortico-subthalamic connection, as the PTN are at rest during the selection task. This is at odd with most computational models of the BG, which assume that CSN and PTN have similar activities, so the STN can perform an Off-Surround effect \citep{Berns94,Mink96,Gurney01a,Girard08} on the GPi/SNr. This simplification is sometimes justified by the idea that the firing activity in the CSN triggers activation of PTN, as they are deeper in the cortical layers. While data in rodents support the existence of a CSN~$\rightarrow$~PTN connection \citep{Morishima06,Kiritani12}, whether it represents a feed-forward activation is still unclear.
Our specific contribution here is to demonstrate that there is no need to have the same signal in the PTN and in the CSN pathways neither to perform selection nor to observe directionality in the STN. We hence suggest that the PTN~$\rightarrow$~STN connection could be serving another purpose, as the {\it braking hypothesis} which states that once a movement is chosen, the excitation of STN by PTN overrides all activity in the GPi/SNr \citep{Gillies04}; conversely, the same overriding mechanism could be applied before selection is ready ("hold your horses", in \cite{Frank06}).

\subsection{Methodology}

\subsubsection{Fixed parameters}
\label{sec:fixparams}


In strict logic, one could (rightly) argue that the fixed parameters are somewhat arbitrary in our study. For example, the neuron numbers are only known to a certain extent, as the comparisons in \cite{Hardman02} show. {\it In fine}, no numerical value is certain for sure, but with approximately 50 free parameters to be set during the optimization process, we are already at the limit of what can be done in a reasonable computational time. Hence, we chose to optimize only the parameters whose interpretation are meaningful (i.e. the connection strength parameters) or for which we have absolutely no reference for adequate values (i.e. the $\theta$ parameters that reflect the neuronal excitability, or the FSI firing rates parameters).

\subsubsection{Optimized but not directly constrained parameters}
\label{sec:fixparams}

Some parameters were optimized, i.e. set by the optimization algorithm, but not present in the construct objective. In particular, the values obtained for the $\theta$ parameters extend to the whole permitted range of [5, 30 mV]. It is hard to compare them to the previous mean-field modeling works of \cite{VanAlbada09a} and \cite{Tsirogiannis10}, given their high heterogeneity (c.f. Table \ref{tab:thetas}) and given the fact that the values in these models were tuned in order to exhibit activity as consistent as possible, under various simplifications which are different from the ones we made. We can however look at their relative values to see how our freely set parameters compare to these fine-tuned values. Alike in these works, $\xyz{\theta}{}{MSN}$ has the highest value while both $\xyz{\theta}{}{GPe}$ and $\xyz{\theta}{}{GPi}$ are set to low values. The comparison of $\xyz{\theta}{}{STN}$ is hard to make, as it is set heterogeneously in previous models: at a middle value for \cite{Tsirogiannis10}, and at a low value for \cite{VanAlbada09a}. Overall, these superficial comparisons show that our optimized $\theta$ parameters are, as far as any parallel can be drawn, globally consistent with previous works, which is remarkable given that there was no direct pressure to set them at any preferred value inside their permitted range.



\subsubsection{Other modeling assumptions}
\label{sec:modsimpl}

Some of the strongest hypotheses in this work stem from the mathematical formalism that we used to describe the dynamics of the neuron populations. In particular, parameters other than the localization of the synapses along the dendrites and the number of such synapses are known to affect the connection strengths - for example, the variations of the number of vesicles in synapses, or the geometrical characteristics of terminals such as their size. Explicitly, the hypotheses that our model formalism convey is that these non-included parameters add, on average, negligible variations of synaptic strength with regards to the variations caused by the number of varicosities and their position on the dendrites. Any elaborate model of the basal ganglia carries such assumptions; however, we propose here an improvement over prior models, as all of our parameters correspond to physical values and are, as such, testable.

\subsection{Related studies}

Computational models of the BG have been helpful to understand the anatomy, physiology and pathophysiology for more than twenty years. Since the first models in the early 90's, there has been two trends in modeling the BG. Works of the first trend used detailed spiking neuron models, trying to be as close as possible to biological data. But the gathering and incorporations of these data is a tremendous task, with a very sharp rise in complexity. These models usually concern only a subpart of the BG, as the hardest parameters to set are the internuclei connection strengths. This trend is illustrated by studies such as detailed modelings of thousands of neurons in the striatum \citep{Wickens07,Humphries09}, or in the GPe $\leftrightarrow$ STN interactions \citep{Kumar11}. 

Works of the second trend use less detailed models, and by doing so, they are able to model the entire BG. Although first models were using rather abstract population models (e.g. leaky integrators in \cite{Berns94,Gurney01a} or lPDS in \cite{Girard08}), more recent ones have been using more plausible models (e.g. fine-tuned spiking neurons in \cite{Humphries06} or mean fields in \cite{VanAlbada09a,Tsirogiannis10}). But the more plausible models have become, the harder they are to tune by hand, resulting after a lot of manual adjustments into one dozen to several dozens of parameters chosen more or less closely to actual biological data.

This dichotomy in models arise from the very essence of modeling, where there are two opposed needs to conciliate. On one hand, there is a need to keep the model as simple as possible, in order to be able to discover and identify general mechanisms. However, the relevance of any model rests on its link to the biological substrate, and is threatened by too much distance to the real BG. Thus any computational model has to balance those two aspect in its conception, which explains the current state with two different classes of models. Our work attempts here to bridge both conceptions by exploring 
the parameters space in a detail-rich model of {\it the whole BG}, while making sure that the models meet {\it quantitative plausibility} criteria.

Previous models adopted the same mathematical formalism (mean-field population model) and were using alternative ways to constrain their parameters. Two recent papers adopted such models \citep{VanAlbada09a,Tsirogiannis10}, and even though these study concerned mainly the oscillations in Parkinson's disease while we focus on the physiology and function of primate BG, both inspired greatly this study.

One similitude in both approaches lies in the setting by hand of internuclei strength parameters during their model design. For both works, this step is backed up by a strong bibliographical review. Unfortunately, this review does not provide the quantitative values they use for most connections, simply because such data is not available. From the theoretical point of view, the succession of arbitrary choices that are then made will ultimately find their justification from the model overall consistency. However, this way of setting parameters has some drawbacks, as different parameter sets could provide such consistency while supporting opposite conclusions on the studied hypothesis.

We transformed this way of setting parameters with "expert knowledge" into two steps: (1) elaboration of wide plausible ranges for both {\it construct} and {\it face} values, backed up by precise references from the literature; and (2) optimization to find parameterizations that are best trades-off for both of these values. Thus, we were able to get rid of the hand-tuning part and moreover to study simultaneously multiple set of parameters instead of one.

Because of this difference in model design, the found strength of parameters are quite different from previous works. The absolute strength of the connections varies, for example, comparing with the values in \cite{VanAlbada09a}, there is a factor of $10^3$ between our optimized values and theirs for the striatofugal connections. The relative strength varies too: for example, GPe~$\rightarrow$~STN connection was set to be roughly three times more potent than STN~$\rightarrow$~GPe in \cite{Tsirogiannis10}, while our study indicates that it should be on the contrary ten times less potent.

To position our methodology among previous studies, it is also relevant to study how other works have used stochastic optimization methods to constrain the parameter choice of BG models. In \cite{Wang07a,Lienard10} the aim of the optimization was to obtain a given functionality, namely selection.
But the neuron models used were very abstract leaky integrators, and their main parameters were synaptic connection strengths coded as a numerical value in an arbitrary range. It is hence quite difficult to link the optimized parameters with the underlying biology. In an inspiring work on BG oscillations, \cite{NevadoHolgado10} modeled the STN and GPe. Their optimization aimed to match the GPe and STN mean firing rates at rest with experimental values, and to reproduce the mean firing rate change of four deactivations by antagonist in the GPe. Their objective is hence very similar to our {\it face objective} except that they ignore the standard deviation in the experimental data fitted, which could possibly be an issue when it is high (e.g. the standard deviation for the fitted gabaergic antagonist in GPe is equal to 80\% of the firing rate). After optimization, they obtained one single parameterization that was capable to fit well to the firing rates of deactivation studies, and with this parameterization they were able to study oscillatory regimes both in the healthy case and in a simulated Parkinson's disease. However, they have no way to assess whether the connection weights correspond plausibly to the neural substrate, hence their method is not suited to study precisely the BG anatomy. Moreover, this method is not scalable to the whole BG because the number of free parameters is then too important compared to the number of published antagonist experiments. Our work here fills these gaps by adding another objective of anatomical plausibility to the optimization.




\begin{acknowledgements}
We warmly thank Dr. Martin Parent for interesting discussion on the single-axon tracing studies, and Dr. Ignasi Cos for helpful comments and suggestions on this work.

\end{acknowledgements}

\section*{Appendix}

\subsection{Mean firing rates at rest}

\label{app:firing_rates_rest}

\notectable{
botcap,
pos = {h},
caption = {Firing rates at rest as reported for some twenty monkeys; the data are represented in the form "Mean $\pm$ SD (number of neurons)". Asterix denote that the firing rate was recorded in the SNr.},%
label = {tab:firing_rates_rest},notespar}{p{1cm}R{2cm}R{2cm}R{2.1cm}}{ 
& \hl{from} \citable{Georgopoulos83a}  & \hl{from} \citable{Mitchell87} &  \hl{from} \citable{Fillion91} \\
  \FL
  STN      & \fr{23}{12}{29}      &                           &                           \\
  GPe      & \fr{71}{31}{87}      &  \fr{56}{21}{89}          &  \fr{76}{28}{108}         \\
  GPi/SNr  & \fr{79}{22}{36}      &  \fr{71}{20}{67}          &  \fr{78}{26}{105}         \\
\\
&  \hl{from} \citable{Matsumura92}            &  \hl{from} \citable{Bergman94}       &  \hl{from} \citable{Bergman94} \\
  \FL
  STN      & \fr{18.5}{12}{265}          &              \fr{17.2}{12}{154}  & \fr{21.6}{9.0}{12}        \\
  GPe      &                             &                                  &                           \\
  GPi/SNr  &                             &              \fr{44.2}{14.8}{16} & \fr{50.5}{13.2}{21}       \\
\\
 &  \hl{from} \citable{Bergman94} &  \hl{from} \citable{Wichmann99}      &  \hl{from} \citable{Wichmann99} \\
  \FL                                                                                       \\
  STN       &  \fr{22.8}{11.8}{54}      &                      &                            \\
  GPe       &                           &                      &                            \\
  GPi/SNr   & \fr{65.0}{15.1}{83}       & \fr{58.5}{1.8}{124}  & \fr{62.4}{3.1}{55}*        \\
\\
 &  \hl{from} \citable{Raz00} &  \hl{from} \citable{Raz00} &  \hl{from} \citable{Kita04} \\
  \FL
STN     &                       &                       & \\
GPe     &  \fr{52}{5.3}{88}     & \fr{73.6}{6}{86}      & \fr{62.6}{25.8}{35} \\
GPi/SNr &  \fr{73.4}{14.8}{8}   & \fr{79.2}{12.7}{21}   & \\
\\
 &  \hl{from} \citable{Kita05}         &  \hl{from} \citable{Starr05} &  \hl{from} \citable{Wichmann06} \\
  \FL
STN     &                         &                         & \fr{23.2}{8.8}{15}         \\
GPe     & \fr{64.1}{24.4}{37}     & \fr{69.7}{3.3}{61}      & \fr{66.1}{19.1}{44}        \\
GPi/SNr & \fr{58.5}{16.8}{29}     & \fr{82.5}{2.5}{96}      & \fr{73.5}{18.1}{67}        \\
\\
 &  \hl{from} \citable{Elias08}          &  \hl{from} \citable{Elias08}&  \hl{from} \citable{Elias08} \\
  \FL
STN       &                         &                        &                         \\
GPe       & \fr{59.8}{16.6}{57}     & \fr{54.6}{15.1}{36}    &                         \\
GPi/SNr   & \fr{69.0}{29.5}{18}     & \fr{74}{13.4}{11}      & \fr{60.5}{18.7}{16}*    \\
\\
          &  \hl{from} \citable{Tachibana08}&  \hl{from} \citable{Iwamuro07} \\
  \FL
STN       &                            & \fr{19}{8.8}{144} \\
GPe       &                            & \\
GPi/SNr   & \fr{71.5}{30.8}{43}        & \\
\LL}

The mean firing rates at rest estimations were derived from the figures in Table \ref{tab:firing_rates_rest}, under the simplifying assumption that each of the monkey recordings $\mu \pm \sigma (n)$ constitute an independent sample of a normally distributed random variable. Using an arbitrary big number $b$, for each set of data we draw $b*n$ random sample from the $\mathcal{N(\mu,\sqrt{\sigma})}$ distribution, and fit the sum of all the simulated data to a normal distribution using least squares.

\subsection{Isoforces of other connections}

\begin{figure}[H]
\begin{tabular*}{\textwidth}{ZZ}
  \includegraphics[width=0.25\textwidth]{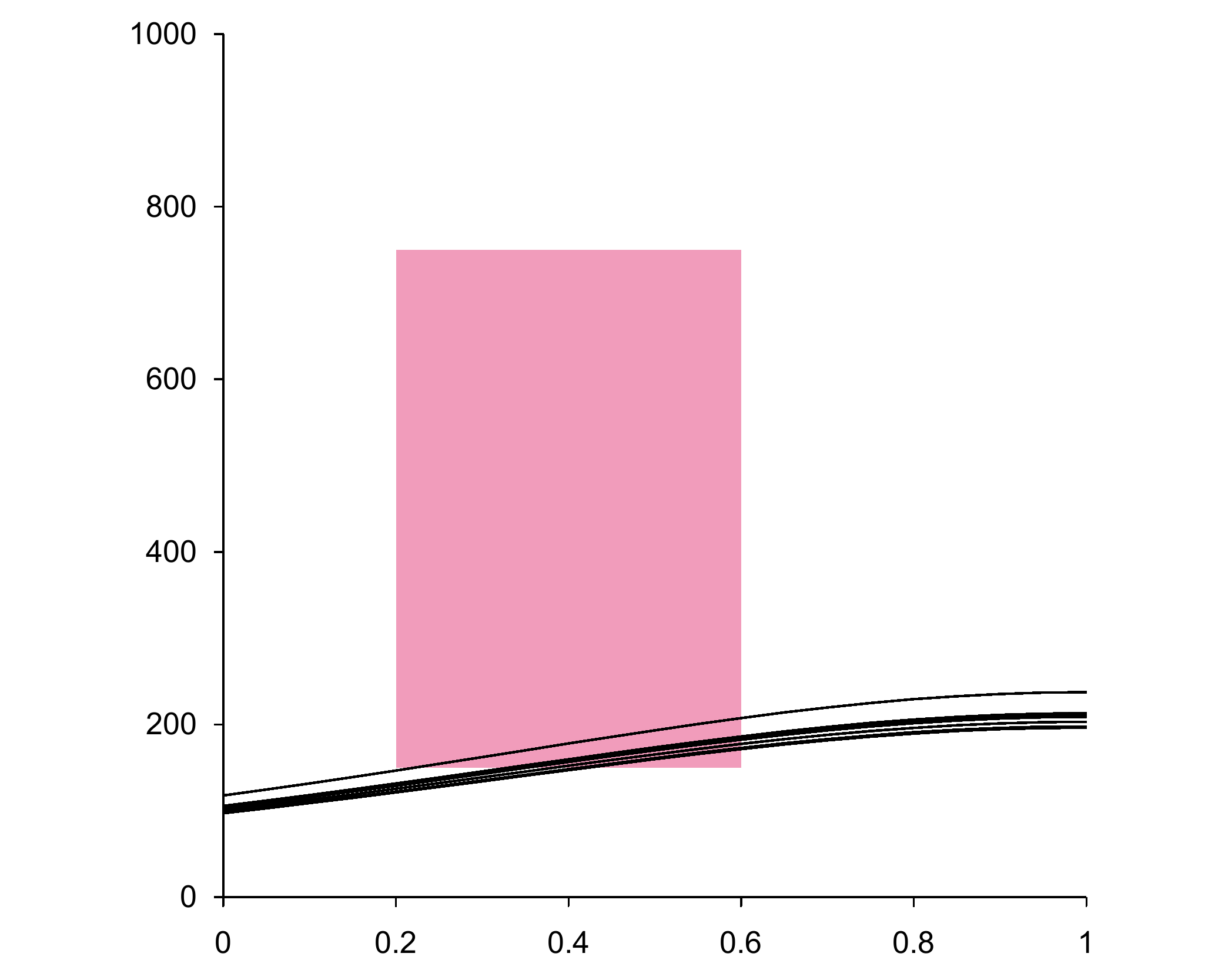} &
  \includegraphics[width=0.25\textwidth]{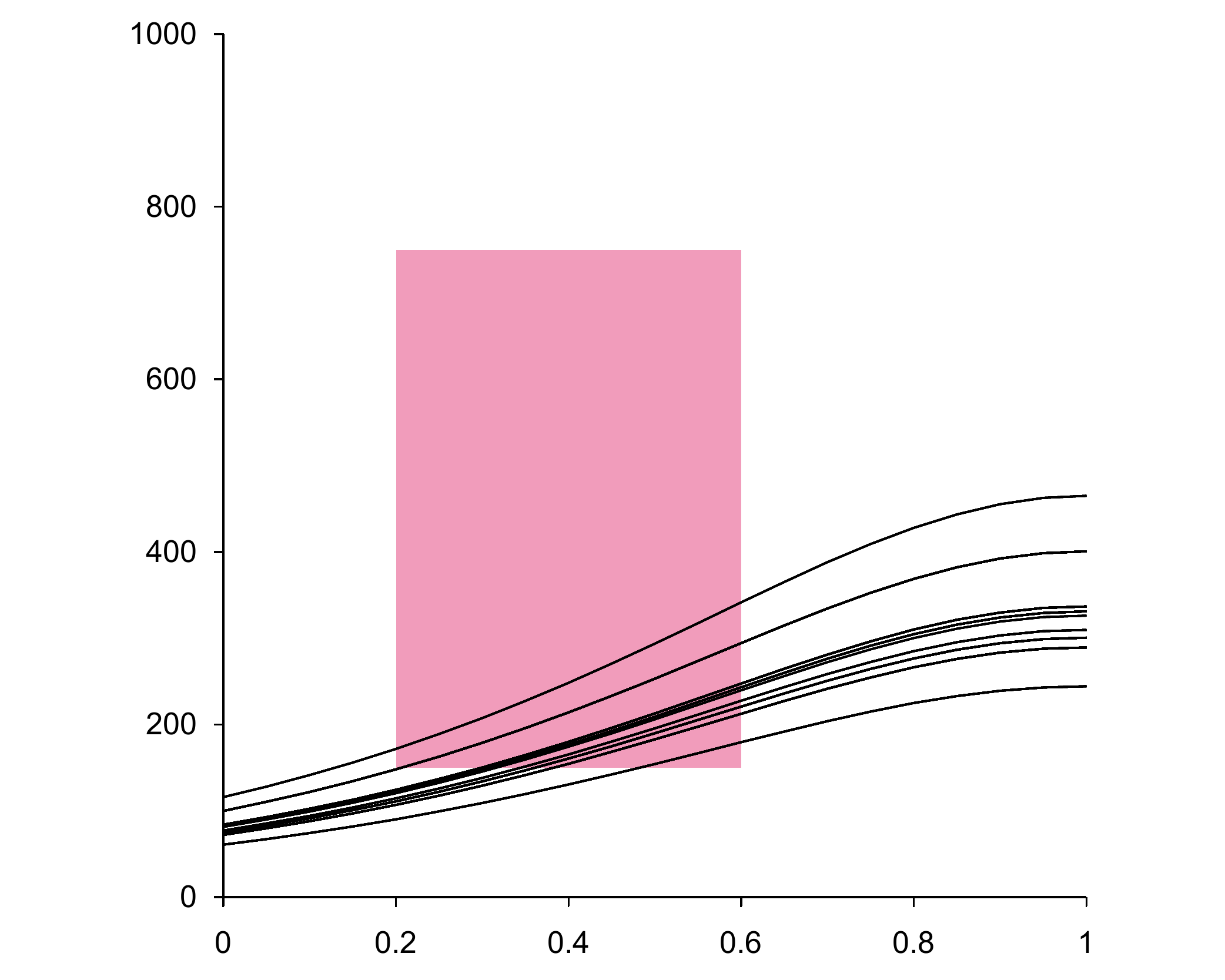}\\
MSN~$\rightarrow$~GPe &
  MSN~$\rightarrow$~GPi/SNr\\
  \includegraphics[width=0.25\textwidth]{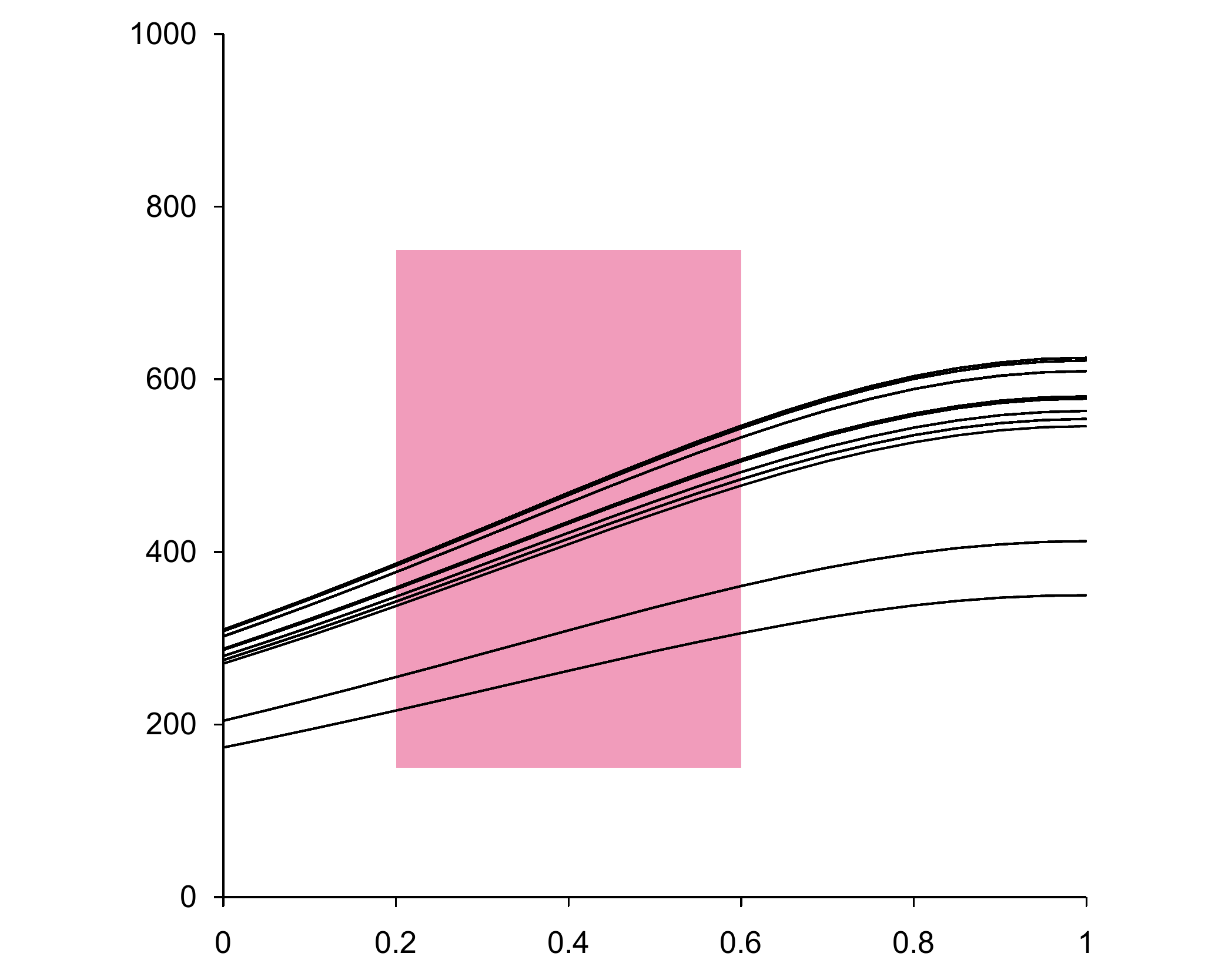} &
  \includegraphics[width=0.25\textwidth]{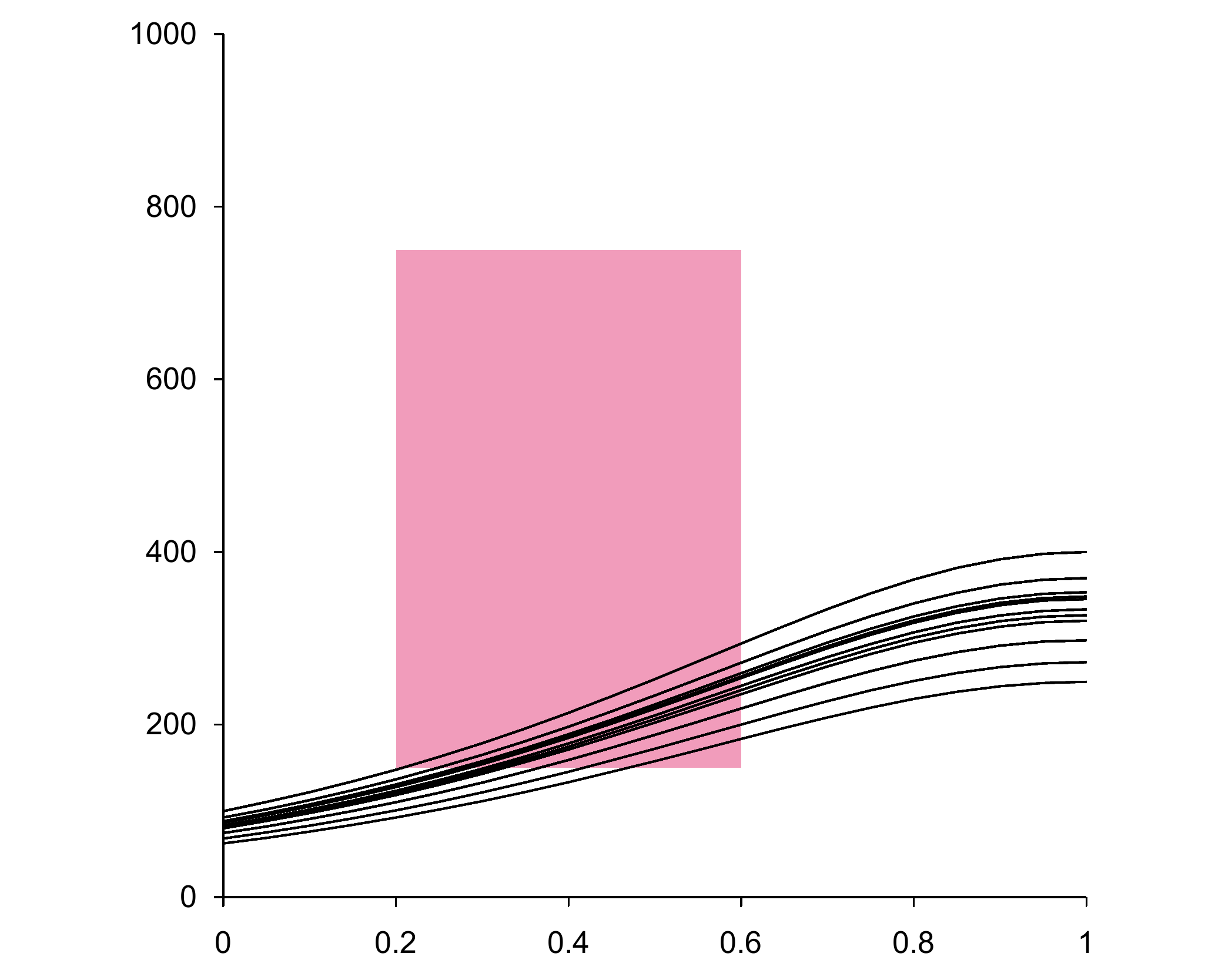}\\
STN~$\rightarrow$~GPe &
   STN~$\rightarrow$~GPi/SNr  \\
  \includegraphics[width=0.25\textwidth]{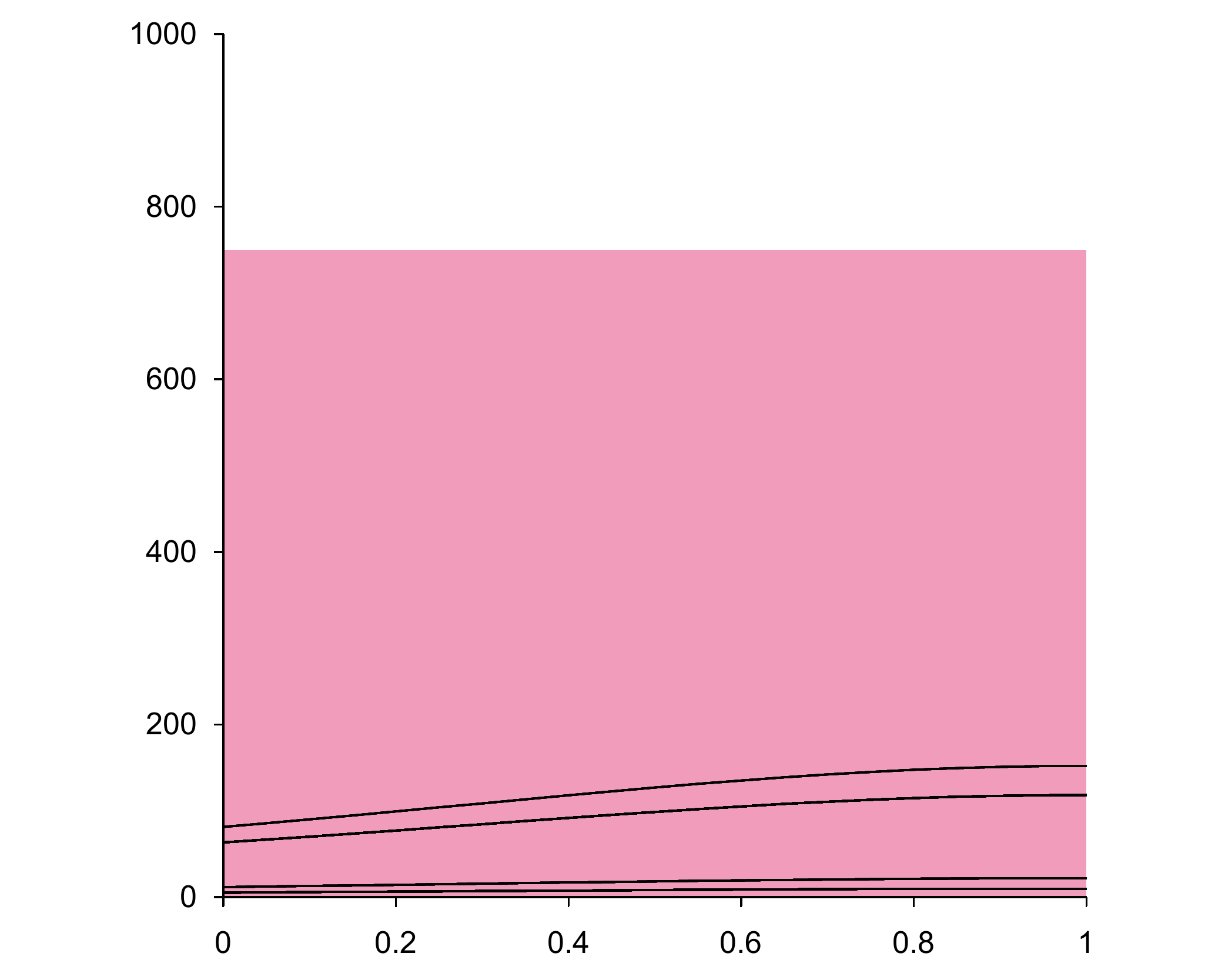} &
  \includegraphics[width=0.25\textwidth]{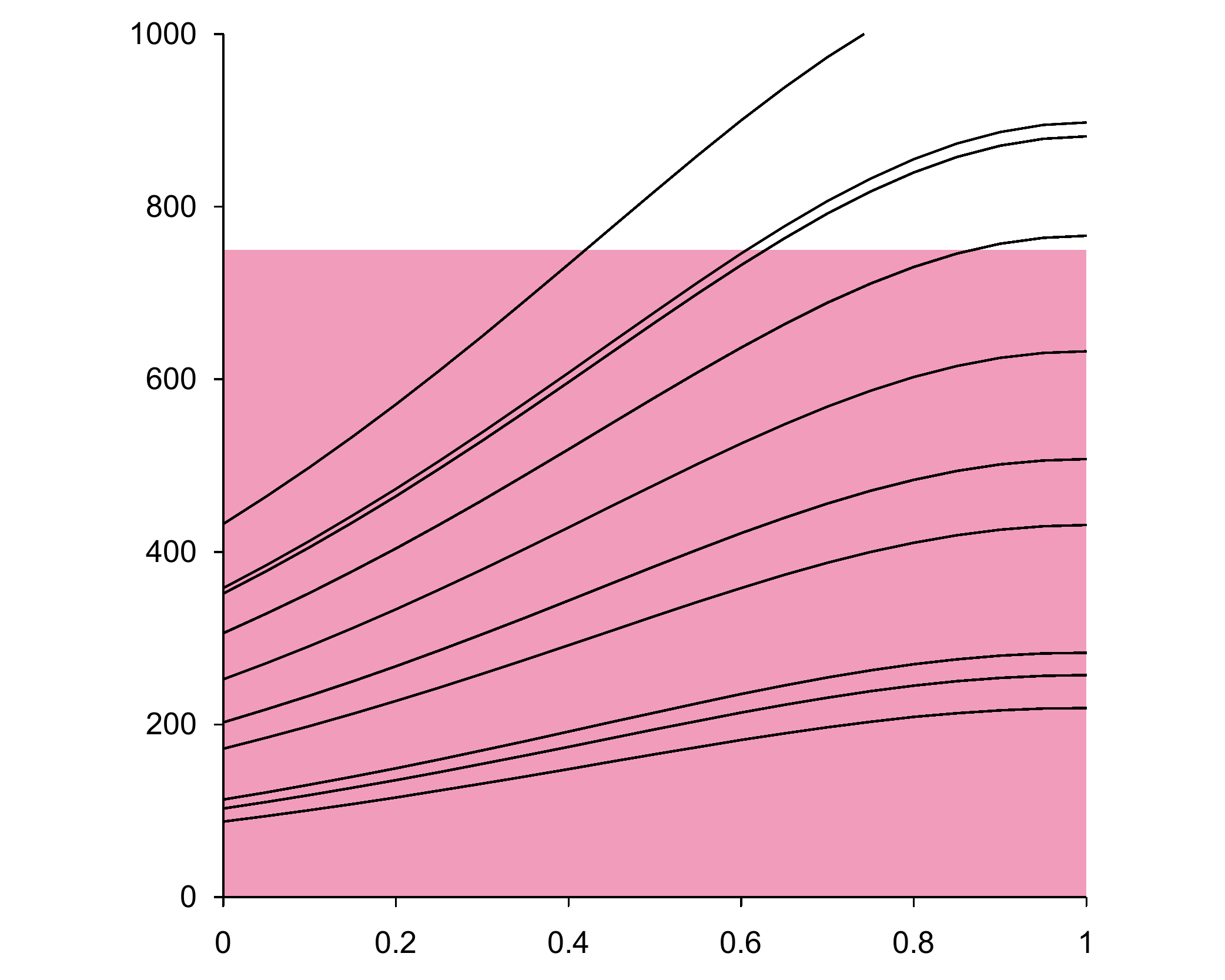}\\
  GPe~$\rightarrow$~MSN &
  GPe~$\rightarrow$~FSI \\
  \includegraphics[width=0.25\textwidth]{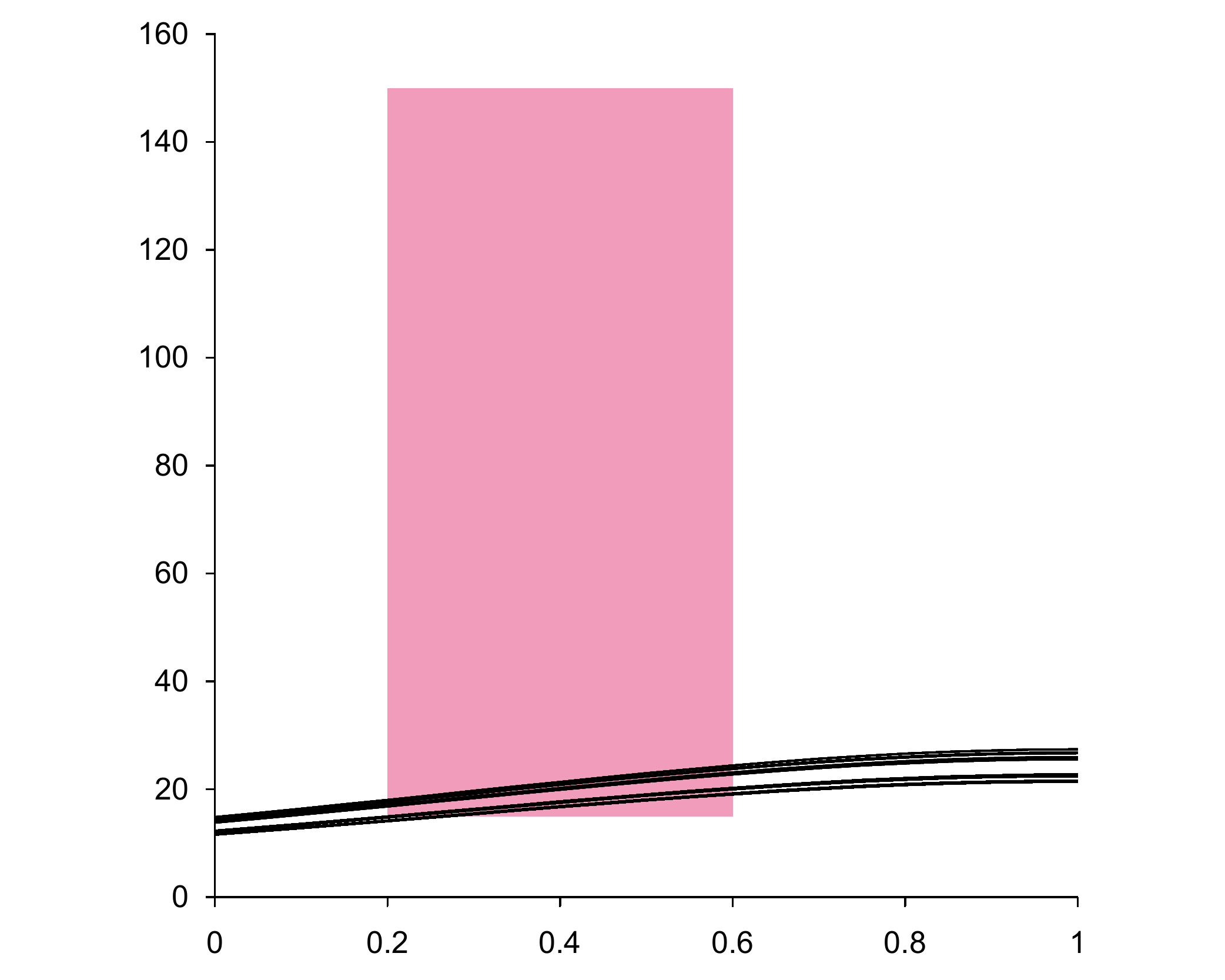}  &
  \includegraphics[width=0.25\textwidth]{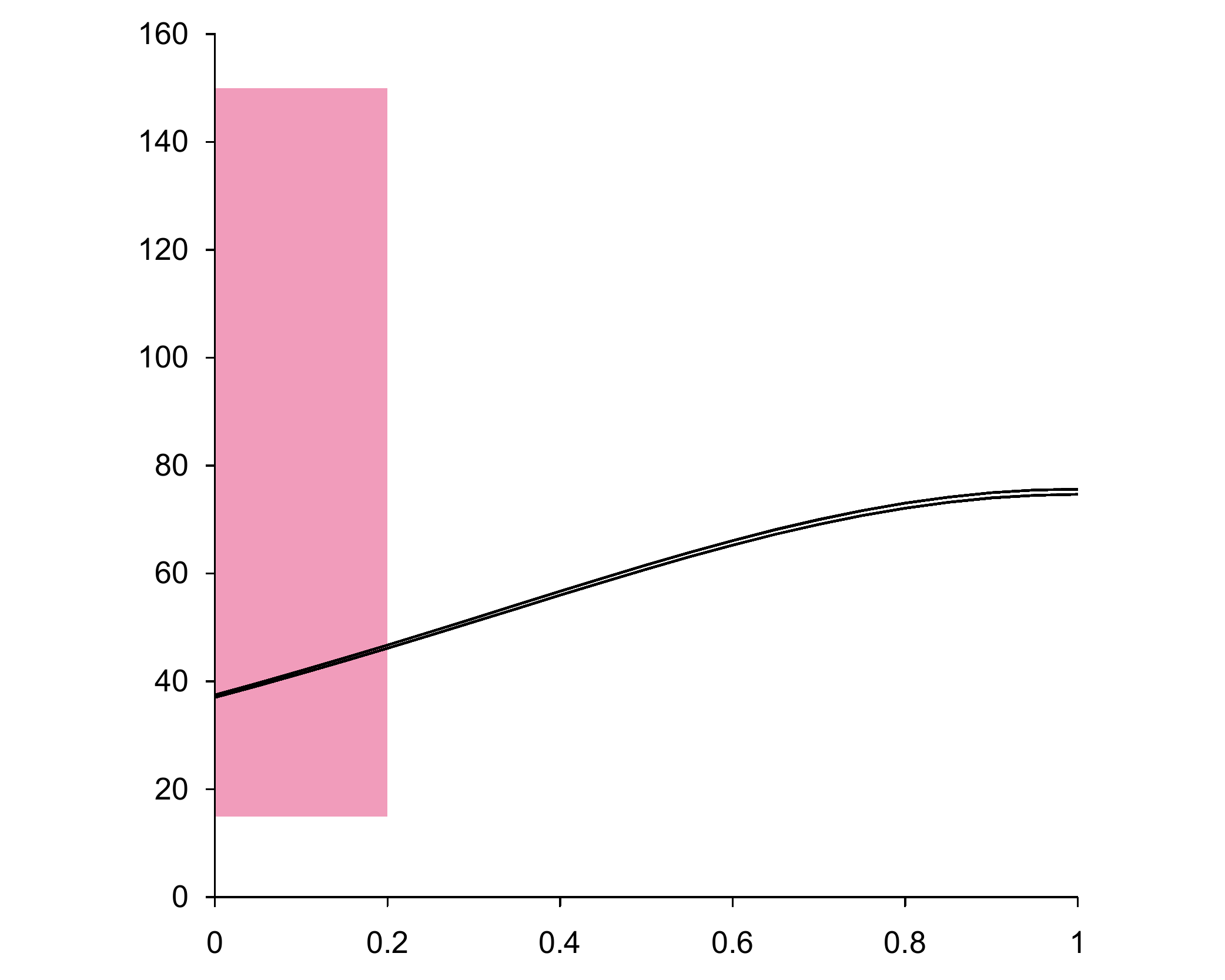} \\
  GPe~$\rightarrow$~STN   &
  GPe~$\rightarrow$~GPe  \\
\end{tabular*}
  \caption{Connection isoforces of other basal ganglia connections discussed in this paper. See the caption of Figure \ref{fig:isoforces} for more explanations.}
  \label{fig:allisoforces}
\end{figure}

As supplementary materials, we provide in Figure \ref{fig:allisoforces} the found isoforces of the MSN and STN projections toward GPe and GPi/SNr, and of the GPe outgoing projections (without the GPe~$\rightarrow$~GPi/SNr projection which is already shown in Figure \ref{fig:isoforces}).

\newcolumntype{Z}{>{\centering\arraybackslash} m{4cm} }

\subsection{Illustration of one solution and source code}

In order to illustrate further the results of our optimization, in Table \ref{tab:exampleparams} are summarized the specific values of the parameters of one solution chosen at random in the pool of optimal solutions. High variations are exhibited in the values of the parameters throughout the set of solutions (as per Tables \ref{tab:minmaxparams} and \ref{tab:thetas}). The exhaustive parameter values of the solutions presented in this paper is available online along with the source code, in order to permit further studies, on the EvoNeuro project website \footnote{http://pages.isir.upmc.fr/EvoNeuro/} and on {\it modeldb} with the accession number 150206 \footnote{http://senselab.med.yale.edu/modeldb/}.

\notectable{
botcap,
mincapwidth = 0.4\textwidth,
caption = {Parameter values for one solution chosen at random among the 1151 different optimal solutions.},%
label = {tab:exampleparams},notespar}{lrlr}{ 
\hl{Parameter} & \hl{Value} & \hl{Parameter} & \hl{Value} \ML
$ \xynu{CSN}{MSN}    $  & 342    &   $ \xyp{CSN}{MSN}  $  & 95 \%  \\
$ \xynu{CSN}{FSI}    $  & 250    &   $ \xyp{CSN}{FSI}  $  & 82 \%  \\
$ \xynu{PTN}{STN}    $  & 259    &   $ \xyp{PTN}{STN}  $  & 97 \%  \\
$ \xyalpha{MSN}{GPe} $  & 171    &   $ \xyp{MSN}{GPe}  $  & 48 \%  \\
$ \xyalpha{MSN}{GPi} $  & 210    &   $ \xyp{MSN}{GPi}  $  & 59 \%  \\
$ \xyalpha{STN}{GPe} $  & 428    &   $ \xyp{STN}{GPe}  $  & 30 \%  \\
$ \xyalpha{STN}{GPi} $  & 233    &   $ \xyp{STN}{GPi}  $  & 59 \%  \\
$ \xyalpha{STN}{MSN} $  & 0      &   $ \xyp{STN}{MSN}  $  & 16 \%  \\
$ \xyalpha{STN}{FSI} $  & 91     &   $ \xyp{STN}{FSI}  $  & 41 \%  \\
$ \xyalpha{GPe}{STN} $  & 19     &   $ \xyp{GPe}{STN}  $  & 58 \%  \\
$ \xyalpha{GPe}{GPi} $  & 16     &   $ \xyp{GPe}{GPi}  $  & 13 \%  \\
$ \xyalpha{GPe}{MSN} $  & 0      &   $ \xyp{GPe}{MSN}  $  & 6  \%  \\
$ \xyalpha{GPe}{FSI} $  & 353    &   $ \xyp{GPe}{FSI}  $  & 58 \%  \\
$ \xyalpha{GPe}{GPe} $  & 38     &   $ \xyp{GPe}{GPe}  $  & 1  \%  \\
$ \xyalpha{FSI}{MSN} $  & 4362   &   $ \xyp{FSI}{MSN}  $  & 19 \%  \\
$ \xyalpha{FSI}{FSI} $  & 116    &   $ \xyp{FSI}{FSI}  $  & 16 \%  \\
$ \xyalpha{MSN}{MSN} $  & 210    &   $ \xyp{MSN}{MSN}  $  & 77 \%  \\
$ \xyalpha{CMPf}{MSN}$  & 4965   &   $ \xyp{CMPf}{MSN} $  & 27 \%  \\
$ \xyalpha{CMPf}{FSI}$  & 1053   &   $ \xyp{CMPf}{FSI} $  & 6  \%  \\
$ \xyalpha{CMPf}{STN}$  & 76     &   $ \xyp{CMPf}{STN} $  & 46 \%  \\
$ \xyalpha{CMPf}{GPe}$  & 79     &   $ \xyp{CMPf}{GPe} $  & 0  \%  \\
$ \xyalpha{CMPf}{GPi}$  & 131    &   $ \xyp{CMPf}{GPi} $  & 48 \%  \\
$ \xynu{PTN}{MSN}    $  & 5      &   $ \xyp{PTN}{MSN}  $  & 98 \%  \\
$ \xynu{PTN}{FSI}    $  & 5      &   $ \xyp{PTN}{FSI}  $  & 70 \%  \\
$ \theta_{\mbox{MSN}} $ & 30 mV    &   $ \theta_{\mbox{GPe}} $ & 11 mV \\
$ \theta_{\mbox{FSI}} $ & 16 mV    &   $ \theta_{\mbox{GPi}} $ & 6 mV  \\
$ \theta_{\mbox{STN}} $ & 26 mV    &   $ \xyFSI{S}{max} $ & 217 Hz     \\
\LL}

\bibliographystyle{apalike}
\bibliography{ref}

\end{document}